\documentclass[twocolumn,floatfix,superscriptaddress,prl]{revtex4-1}
\usepackage{bm,graphicx,url,epsf,color,amssymb}
\usepackage[french]{babel}

\begin{document}

\title{Tunable Quantum Criticality and Super-ballistic Transport in a `Charge' Kondo Circuit}

\author{Z. Iftikhar}
\affiliation{Centre de Nanosciences et de Nanotechnologies (C2N), CNRS, Univ Paris Sud, Universit\'e Paris-Saclay, 91120 Palaiseau, France}
\author{A. Anthore}
\affiliation{Centre de Nanosciences et de Nanotechnologies (C2N), CNRS, Univ Paris Sud, Universit\'e Paris-Saclay, 91120 Palaiseau, France}
\affiliation{Univ Paris Diderot, Sorbonne Paris Cit\'e, 75013 Paris, France}
\author{A.K. Mitchell}
\affiliation{School of Physics, University College Dublin, Dublin 4, Ireland}
\author{F.D. Parmentier}
\affiliation{Centre de Nanosciences et de Nanotechnologies (C2N), CNRS, Univ Paris Sud, Universit\'e Paris-Saclay, 91120 Palaiseau, France}
\author{U. Gennser}
\affiliation{Centre de Nanosciences et de Nanotechnologies (C2N), CNRS, Univ Paris Sud, Universit\'e Paris-Saclay, 91120 Palaiseau, France}
\author{A. Ouerghi}
\affiliation{Centre de Nanosciences et de Nanotechnologies (C2N), CNRS, Univ Paris Sud, Universit\'e Paris-Saclay, 91120 Palaiseau, France}
\author{A. Cavanna}
\affiliation{Centre de Nanosciences et de Nanotechnologies (C2N), CNRS, Univ Paris Sud, Universit\'e Paris-Saclay, 91120 Palaiseau, France}
\author{C. Mora}
\affiliation{Laboratoire Pierre Aigrain, Ecole Normale Sup\'{e}rieure-PSL Research University, CNRS, UPMC Univ Paris 06-Sorbonne Universit\'es, Univ Paris Diderot-Sorbonne Paris Cit\'e, 75005 Paris, France}
\author{P. Simon}
\affiliation{Laboratoire de Physique des Solides, CNRS, Univ. Paris-Sud, Universit\'e Paris-Saclay, 91405 Orsay, France}
\author{F. Pierre}
\email{frederic.pierre@u-psud.fr}
\affiliation{Centre de Nanosciences et de Nanotechnologies (C2N), CNRS, Univ Paris Sud, Universit\'e Paris-Saclay, 91120 Palaiseau, France}

\begin{abstract}
Quantum phase transitions are ubiquitous in many exotic behaviors of strongly-correlated materials. 
However the microscopic complexity impedes their quantitative understanding. 
Here, we observe thoroughly and comprehend the rich strongly-correlated physics in two profoundly dissimilar regimes of quantum criticality.
With a circuit implementing a quantum simulator for the three-channel Kondo model, we reveal the universal scalings toward different low-temperature fixed points and along the multiple crossovers from quantum criticality. 
Notably, an unanticipated violation of the maximum conductance for ballistic free electrons is uncovered.
The present charge pseudospin implementation of a Kondo impurity opens access to a broad variety of strongly-correlated phenomena.
\end{abstract}

\maketitle

Quantum phase transitions (QPT) are believed to underpin many intriguing quantum states of matter and unconventional behaviors \cite{Sachdev2011}.
Although they take place at absolute zero temperature while precisely tuning a control parameter, such as the magnetic field, continuous QPTs of second order are accompanied by the development of a highly correlated quantum critical state that extends over a broadening range of parameters as temperature is increased.
This regime of quantum criticality provides a universal description of very diverse strongly-correlated systems whose properties obey scaling laws according to the QPT universality class, and not to microscopic details.
While QPTs are ubiquitous in contemporary theoretical physics, the challenge is to find well-controlled experimental systems for their exploration and quantitative comprehension. 

Tunable nanostructures provide a path to a microscopic understanding that circumvents the complexity of real-world highly correlated materials.
So far, however, the rare examples that exhibit a second-order QPT \cite{Potok2007,Mebrahtu2012,Mebrahtu2013,Keller2015,Iftikhar2015} demonstrate only a single quantum critical point (associated with the two-channel Kondo effect, described below).
Although non-Fermi liquid, this critical point can be reduced to a non-interacting system allowing for perturbative approaches at low temperatures \cite{Emery1992,Coleman1995}.
Here, a completely characterized circuit embodies the three-channel Kondo model, with three fully tunable channels connected to a magnetic impurity emulated by the charge states of a metallic island.
This gives us access with the same nanostructure to two universality classes of quantum criticality (connected with the two-channel and three-channel Kondo effects) that manifest profoundly dissimilar physics.
For instance, the quantum critical point for two symmetric Kondo channels can be understood in terms of free electrons and Majorana fermions \cite{Emery1992,Coleman1995}, whereas for three symmetric channels it involves ($\mathbb{Z}_3$) parafermions in irreducibly strong interactions \cite{Affleck2001}.
The demonstrated high-precision implementation qualifies our device as an analogue quantum simulator, providing quantitative experimental solutions for the three-channel Kondo model.

The main findings further detailed in the following sections are now briefly summarized.

We first explore the two-channel Kondo (2CK) and three-channel Kondo (3CK) quantum critical physics, by precisely tuning the connected Kondo channels at symmetry.
The different low temperature convergence (fixed) points are determined for the channels' conductance, and a perfect quantitative agreement with the universal values predicted theoretically is found.
This constitutes a direct experimental signature of the Kondo coupling renormalization flow toward an intermediate fixed point.
The 2CK and 3CK non-Fermi liquid character is also established, through their distinctive temperature power-laws.
More generally, the full conductance renormalization flow is unveiled, from asymptotic freedom to quantum criticality, and confronted to state-of-the-art numerical renormalization group (NRG) calculations.
In addition, a connection is established between experimental scaling Kondo temperature and microscopic model parameters.

Secondly, we determine the range of parameters where quantum criticality applies (different for 2CK and 3CK) and explore the many-body physics that develops as the system flows away from quantum criticality upon reducing temperature.
For this purpose, the device is controllably detuned from the symmetric 2CK and 3CK quantum critical points.
By breaking the Kondo impurity degeneracy, we observe the generically expected universal scaling character of the crossover from quantum criticality.
The crossover scaling temperature experimentally extracted is found to increase as a power law of degeneracy breaking that closely verifies the corresponding 2CK or 3CK predictions.
Moreover, the measured full universal conductance scaling curves precisely match with theoretical calculations (analytical at 2CK, NRG at 3CK).
By breaking the symmetry between channels instead of the Kondo impurity degeneracy, the full experimental renormalization flow of three channels competing to screen the Kondo impurity is plainly exposed.
Remarkably, we observe that the conductance across one channel can markedly exceed the maximum quantum limit for free electrons, as corroborated by new NRG calculations.\\

\begin{figure}[!tbh]
\renewcommand{\figurename}{Fig.}
\centering
\includegraphics[width=\columnwidth]{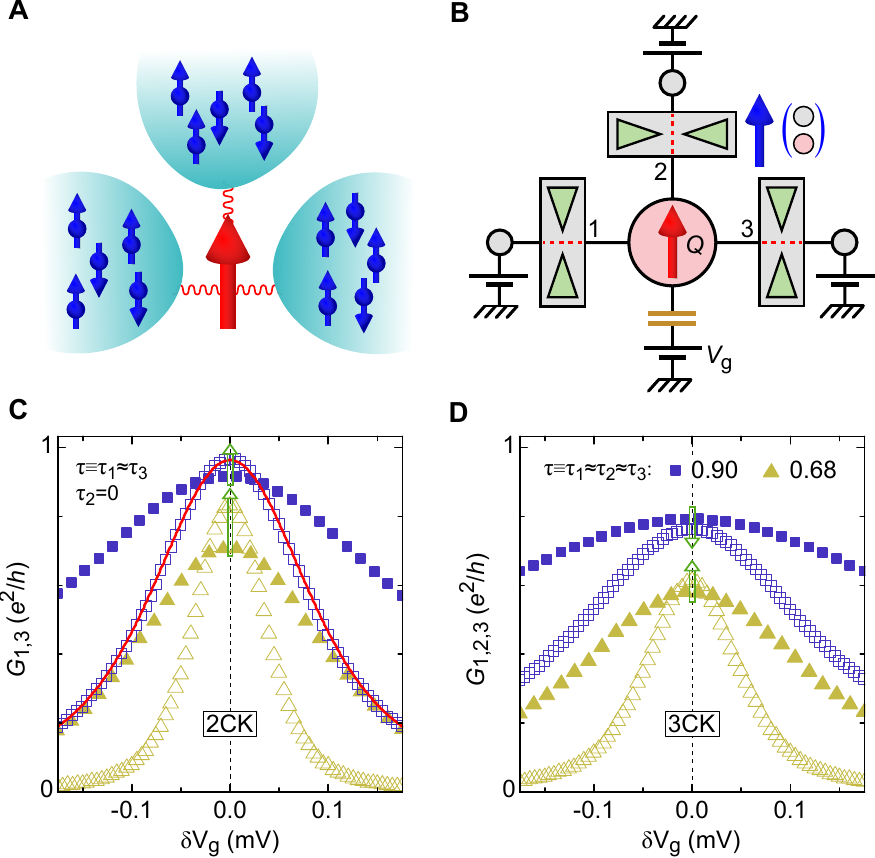}
\caption{
\footnotesize
\textbf{Multi-channel Kondo model and `charge' implementation}. 
\textbf{A}, In the Kondo model, a local spin (red arrow) is antiferromagnetically coupled to the spin of electrons (blue arrows). 
Each Kondo channel corresponds to one distinct electron continuum (here three). 
\textbf{B}, Sample schematic realizing the `charge' pseudospin implementation of the three-channel Kondo model.
A micrometer-scale metallic island (red disk) is connected to large electrodes (small gray disks) through three QPCs (green split gates), each set to a single (spin-polarized) conduction channel (red dashed lines) indexed by $i\in\{1,2,3\}$.
\textbf{C, D}, Quantum channels conductance measured versus gate voltage $V_\mathrm{g}$ are displayed over half a Coulomb oscillation period $\Delta\simeq0.7$\,mV (several sweeps including different consecutive peaks are averaged). 
Measurements at $T\simeq7.9$\,mK and 29\,mK are shown, respectively, as open and full symbols for two (C) or three (D) symmetric channels.
The squares (triangles) correspond to an `intrinsic', unrenormalized transmission probability across the connected QPCs of $\tau\simeq0.90$ ($\tau\simeq0.68$).
The red continuous line (C) displays the $T=7.9$\,mK prediction for two channels both set to $\tau=0.90$ \cite{supplementary}.
Green arrows indicate the conductance evolution at $\delta V_\mathrm{g}=0$ as temperature is reduced. 
}
\label{Fig_NCK}
\end{figure}

{\noindent\textbf{The multi-channel Kondo model}.}\\
The multi-channel Kondo model, which generalizes the original (one channel) Kondo model, gives rise to archetypal QPTs and collective, non-Fermi liquid behaviors from a minimal Hamiltonian.
Although introduced to account for the different atomic orbitals in metals \cite{Nozieres1980,Hewson1997,Cox1998}, it has developed over the years into a central testing ground for strongly-correlated and quantum critical physics, and a benchmark for many-body theoretical methods \cite{Emery1992,Affleck1993,Matveev1995,Hewson1997,Cox1998,Pustilnik2004b,Vojta2006,Bulla2008,Sela2011,Mitchell2016}.
The model describes a local Kondo spin $S$ (here $1/2$) coupled antiferromagnetically to $N$ independent free-electron continua ($N=3$ in Fig.~1A):
\begin{equation}
H_{\mathrm{NCK}}=\sum_{i=1}^N J_i~\mathbf{s_i}\cdot\mathbf{S}+H_\mathrm{continua},\label{HNCK}
\end{equation}
with $H_{\mathrm{NCK}}$ the $N$-channel Kondo Hamiltonian, $\mathbf{s_i}$ the local spin density of electron continuum (channel) $i$ at the Kondo spin $\mathbf{S}$ location, $J_i>0$ the coupling strengths (here isotropic) and $H_\mathrm{continua}$ the free-electron continua Hamiltonian.
The conventional single-channel model ($N=1$) exhibits universal scaling, but no second-order QPT or non-Fermi liquid physics.
As the temperature $T$ is reduced, the electrons progressively screen the Kondo spin, resulting for $T\rightarrow0$ in an idle spin-singlet \cite{Hewson1997}.
In contrast, for $N\geq2$, there is a competition between channels to screen the $S=1/2$ Kondo impurity, which develops into second-order QPTs.
Each number of identical channels corresponds to a different class of quantum criticality \cite{Vojta2006}, with specific non-Fermi liquid physics \cite{Cox1998} and collective excitations revealed by e.g. a divergent specific heat coefficient $c/T$ as $T\rightarrow0$.
The marginal two-channel case corresponds to a logarithmic $c/T$ divergence \cite{Cox1998} while power law $c/T$ divergences are predicted for $N\geq3$ \cite{Cox1998}.\\

{\noindent\textbf{Kondo `charge' pseudospin implementation}.}\\
Experimentally, Kondo nanostructures are usually small quantum dots \cite{Goldhaber-Gordon1998a,Cronenwett1998,Nygard2000,Buitelaar2002} where coherent electron cotunneling merges the distinct electrical contacts into one Kondo channel \cite{Glazman1988,Ng1988} (except in two-channel devices exploiting Coulomb blockade to suppress cotunneling \cite{Oreg2003,Florens2007,Potok2007,Keller2015}).
In contrast, the recently demonstrated \cite{Iftikhar2015} `charge' Kondo approach \cite{Matveev1991,Matveev1995,Furusaki1995b} is here extended to three independent Kondo channels.
As illustrated in Fig.~1B, the `charge' Kondo impurity $\mathbf{S}$ is not a magnetic spin, but a pseudospin-$1/2$ (red arrow) built from the macroscopic quantum states describing the overall charge $Q$ of a small metallic island (red disk).
In the most straightforward case of a weakly connected island whose charge is well quantized \cite{Jezouin2016}, the Kondo spin $S=\{\downarrow,\uparrow\}$ directly maps on the island's two charge states of lowest energy $\{Q,Q+e\}$.
All the other charge configurations are indeed frozen-out and can be ignored at low temperatures $T\ll E_\mathrm{C}/k_\mathrm{B}$ ($E_\mathrm{C}=e^2/2C$ the charging energy, $e$ the electron charge, $C$ the island geometric capacitance, $k_\mathrm{B}$ the Boltzmann constant).
The `charge' pseudospin energy degeneracy is obtained by tuning with a gate voltage $V_\mathrm{g}$ the device at the degeneracy point between the charge states $Q$ and $Q+e$.
Note that detuning $V_\mathrm{g}$ away from charge degeneracy is completely analogous to applying a magnetic field on usual magnetic Kondo impurities \cite{Matveev1991}.
The island charge Kondo pseudospin $S$ is however not coupled to the real spin of electrons.
Instead, it is flipped by transferring electrons in and out of the island, through the connected electrical channels (red dashed lines).
By labeling the location of each electronic states along a channel, this mechanism takes the form of a Kondo (pseudo)spin-exchange coupling:
Introducing an electron pseudospin $s$ (blue arrow), which corresponds to the electron localization inside ($s=\downarrow$) or outside ($s=\uparrow$) of the island, the tunneling of an electron flips both its localization pseudospin $s$ as well as the island overall charge pseudospin $S$.
Note that a well-developed Kondo effect requires a continuum of electronic states for both localization pseudospins.
This implies a continuous density of states in the metallic island (in contrast with small quantum dots).
Importantly, the different conduction channels constitute here separate Kondo channels.
Note also that the same physics is predicted at $T\ll E_\mathrm{C}/k_\mathrm{B}$ for arbitrary connection strengths \cite{Matveev1995,Lebanon2003,supplementary}, except perfectly ballistic contacts, despite the coexistence of many charge states in a quantum superposition near the ballistic limit \cite{Jezouin2016}.
In practice, we find from NRG calculations performed with a broad range of $\tau$ that $T\lesssim E_\mathrm{C}/20k_\mathrm{B}$ ensures negligible deviations from universal Kondo physics \cite{supplementary}.

Each of the Kondo/conduction channels passes through a different quantum point contact (QPC) individually formed and tuned by field effect in a high-mobility Ga(Al)As two-dimensional electron gas \cite{supplementary}. 
Single channels, polarized in real electron spin, are obtained by immersing the device into a large magnetic field ($B\simeq2.7$\,T, corresponding to the regime of the integer quantum Hall effect at filling factor $\nu=3$).
The Kondo channel couplings $J_\mathrm{i}$ ($i\in\{1,2,3\}$) are individually characterized by the `intrinsic' (i.e. unrenormalized by Kondo or Coulomb effects) transmission probability $\tau_\mathrm{i}$ across the unique open transport channel of QPC$_\mathrm{i}$. 
The micron-scale separation between QPCs enables independent fine-tuning ($\sim0.1\%$) and high-precision characterization ($\lesssim2\%$, with a large dc bias voltage suppressing Kondo/Coulomb renormalization \cite{supplementary}) of the Kondo channels, over the full range $\tau_\mathrm{i}\in[0,1]$.
Such fine-tuning of the connected channels to identical couplings is crucial to approach the frustrated, symmetric Kondo critical points.
The two-channel Kondo (2CK) configurations are implemented by setting $\tau_1\simeq\tau_3\equiv\tau$ and $\tau_2=0$, whereas for the three-channel Kondo (3CK) configurations $\tau_1\simeq\tau_2\simeq\tau_3\equiv\tau$.
With the charging energy $E_\mathrm{C}\simeq k_\mathrm{B}\times0.3$\,K (separately obtained from Coulomb diamond measurements) and high-precision shot-noise thermometry \cite{Iftikhar2016}, the device is completely characterized.
The knowledge of these parameters indeed allows for a full quantitative microscopic understanding \cite{Matveev1991,Furusaki1995b,Mitchell2016}.
In practice, Kondo physics is observed through the renormalized QPC conductances $G_i$ measured in-situ.
As the symmetry between channels is found preserved by renormalization (at an experimental accuracy of $\sim0.003e^2/h$), we generally display the averages $G_{1,3}\equiv(G_1+G_3)/2$ and $G_{1,2,3}\equiv(G_1+G_2+G_3)/3$ when investigating the symmetric 2CK and 3CK configurations, respectively.

Figure~1C,D validates the high-precision implementation/quantum simulation of the `charge' Kondo model, and qualitatively illustrates the different two-channel (C) and three-channel (D) Kondo behaviors.
The renormalized conductance across channels tuned to `intrinsic' $\tau\simeq0.90$ (squares) or $0.68$ (triangles) is displayed for $T\simeq7.9$\,mK and $29$\,mK while sweeping the gate voltage $V_\mathrm{g}$.
The charge degeneracy point is identified as the conductance peak ($\delta V_\mathrm{g}=0$).
The perfect match, without any fit parameters, between the conductance data and the quantitative predictions of the `charge' Kondo model derived analytically for two near ballistic channels at low temperature (continuous line in Fig.~1C, \cite{Furusaki1995b,supplementary}) attests to the accurate device characterization and to its precise implementation of the model for arbitrary  Kondo pseudospin energy splitting (see also \cite{Mitchell2016,supplementary}).
At large $\delta V_\mathrm{g}$, the conductance is systematically reduced upon lowering $T$ as usually expected from plain charge quantization.
At $\delta V_\mathrm{g}=0$ and for two or three symmetric channels set to $\tau\simeq0.68$, we observe instead a conductance increase due to the Kondo renormalization of weakly connected channels.
Setting the device at the larger $\tau\simeq0.90$ results in qualitatively different conductance renormalizations at $\delta V_\mathrm{g}=0$ for 2CK and 3CK, with opposite signs.\\

\begin{figure}[!t]
\renewcommand{\figurename}{Fig.}
\centering
\includegraphics[width=\columnwidth]{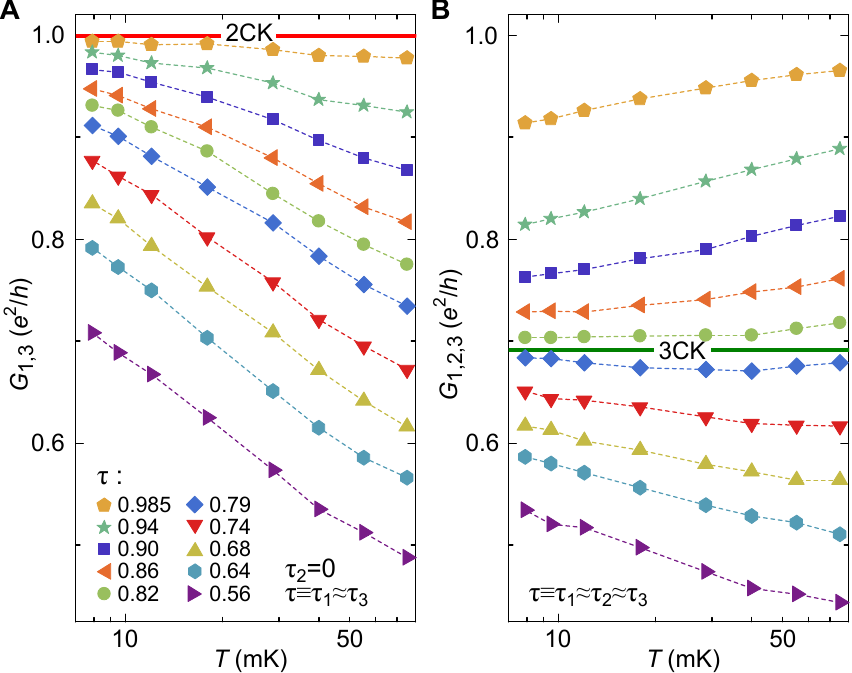}
\caption{
\footnotesize
\textbf{Quantum critical fixed points}. 
The conductance of two (A) or three (B) symmetric channels measured at the charge degeneracy point ($\delta V_\mathrm{g}=0$) is plotted as symbols versus temperature on a logarithmic scale.
Each set of identical symbols connected by dashed lines corresponds to the same device setting ($\tau$).
The predicted 2CK (A) and 3CK (B) low temperature fixed points for the conductance per channel in the present `charge' Kondo implementation are shown as horizontal continuous lines ($G_\mathrm{2CK}=e^2/h$, $G_\mathrm{3CK}=2\sin^2(\pi/5)e^2/h$).
}
\label{Fig_fixedpoints}
\end{figure}

{\noindent\textbf{Observation of an intermediate non-trivial fixed point.}}\\
The above findings corroborate the theoretical expectations for the different 2CK and 3CK low-temperature conductance fixed points \cite{Furusaki1995b,Yi1998}.
Both 2CK and 3CK quantum critical fixed points are associated with an intermediate value of the renormalized Kondo coupling $0<|J|<\infty$ \cite{Nozieres1980,Cox1998}.
In previous experiments on small quantum dots \cite{Potok2007,Keller2015}, the 2CK intermediate coupling could not be established.
Indeed, $T$ was not low enough with respect to the scaling Kondo temperature $T_\mathrm{K}$ to show a saturation and, furthermore, a trivial intermediate asymptotic value of the measured conductance is also generated by asymmetries between electrical channels \cite{Pustilnik2004b}, and therefore does not itself imply an intermediate coupling in these spin Kondo devices.
Moreover, the intermediate coupling character of the 2CK fixed point is not entirely invariable, but depends on the choice of representation \cite{Emery1992,Coleman1995,Matveev1995,Furusaki1995b}.
In particular, the 2CK fixed point can be described as a non-interacting system involving two Majorana modes (one free, one in the strong coupling limit \cite{Emery1992}).
This dual strong-coupling character of the 2CK fixed point also materializes in the present `charge' Kondo implementation: 
Here $G_{1,3}$ constitutes an alternative probe of the coupling between electrons and `charge' Kondo impurity, which flows not toward an intermediate value per electrical channel but toward the maximum free-electron quantum limit $G_\mathrm{2CK}=e^2/h$ \cite{Furusaki1995b}.
In contrast, the genuinely intermediate character of the interacting 3CK fixed point is predicted to show up directly in `charge' Kondo circuits, as a flow of the conductance per channel $G_{1,2,3}$ toward the non-trivial intermediate universal conductance $G_\mathrm{3CK}=2\sin^2(\pi/5)e^2/h\simeq0.691e^2/h$ \cite{Yi1998}.

The precise 2CK and 3CK low-temperature universal conductance fixed points are experimentally established by measuring the temperature evolution of $G_{1(,2),3}$ for a broad range of symmetric channel settings ($\tau\in[0.56,0.985]$). 
For this purpose, and until explicitly specified, the device is tuned at charge degeneracy ($\delta V_\mathrm{g}=0$) where Kondo effect is expected.
Figure~2 displays measurements of $G_{1(,2),3}$ as symbols, versus $T$ in logarithmic scale.
In the 2CK configuration (A), whatever the setting $\tau$, we find that $G_{1,3}$ always grows as $T$ is reduced.
This observation validates the predicted $e^2/h$ Kondo fixed point (horizontal red line), at an experimental accuracy of $0.006e^2/h$ (see also \cite{Iftikhar2015}).
Upon lowering $T$ in the 3CK configuration (B), $G_{1,2,3}$ systematically grows when below $0.68e^2/h$ ($T\leq40$\,mK), and decreases when above $0.70e^2/h$.
This validates the predicted 3CK universal conductance fixed point $G_\mathrm{3CK}\simeq0.69e^2/h$ (horizontal green line) at an experimental accuracy of $\pm0.01e^2/h$.
This constitutes a direct experimental evidence of an intermediate non-Fermi liquid fixed point.\\

\begin{figure}[!t]
\renewcommand{\figurename}{Fig.}
\centering
\includegraphics[width=\columnwidth]{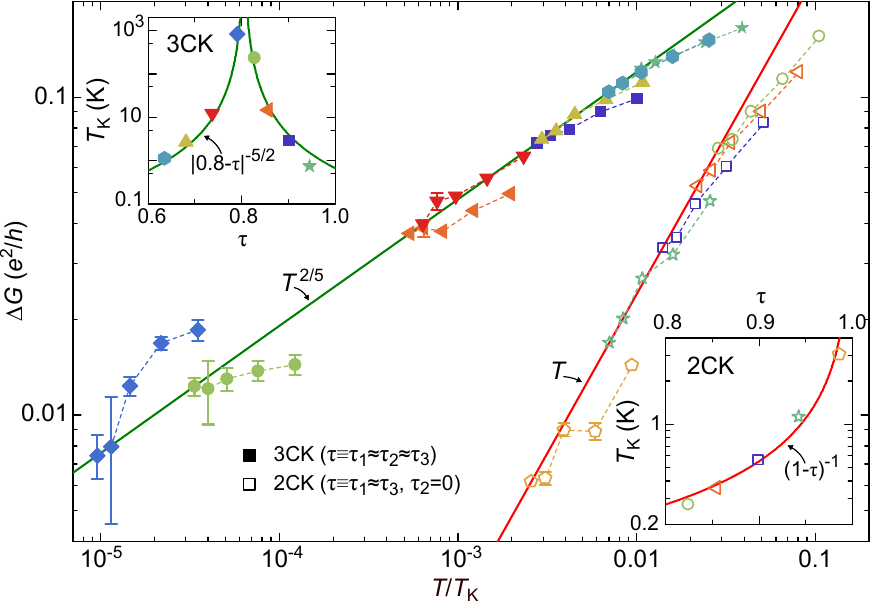}
\caption{
\footnotesize
\textbf{Non-Fermi liquid scaling exponents}. 
The absolute difference between symmetric channels conductance at charge degeneracy and predicted Kondo fixed point ($\Delta G\equiv|G_{1(,2),3}-G_\mathrm{2CK(3CK)}|$) is plotted as symbols (open/full for 2CK/3CK) versus $T/T_\mathrm{K}$ in a $\log$-$\log$ scale for $T\in\{7.9$,$9.5$,$12$,$18$,$29\}$\,mK.
Statistical error bars are shown when larger than symbols.
The red and green continuous straight lines display the predicted power-law scaling at $T/T_\mathrm{K}\ll1$ for the conductance per channel in the present `charge' 2CK and 3CK implementations, respectively.
The scaling Kondo temperature $T_\mathrm{K}$ is adjusted separately for each tuning $\tau$ of the symmetric channels (see corresponding symbols in insets).
This is done by matching the lowest temperature data point $\Delta G(T\simeq7.9$\,mK) with the corresponding displayed power-law.
Continuous lines in insets show the predicted power-law divergences of $T_\mathrm{K}$ versus $\tau$ for 2CK (bottom right inset) and 3CK (top left inset).
}
\label{Fig_powerlaw}
\end{figure}

\begin{figure*}[!tbh]
\renewcommand{\figurename}{Fig.}
\centering
\includegraphics[width=180mm]{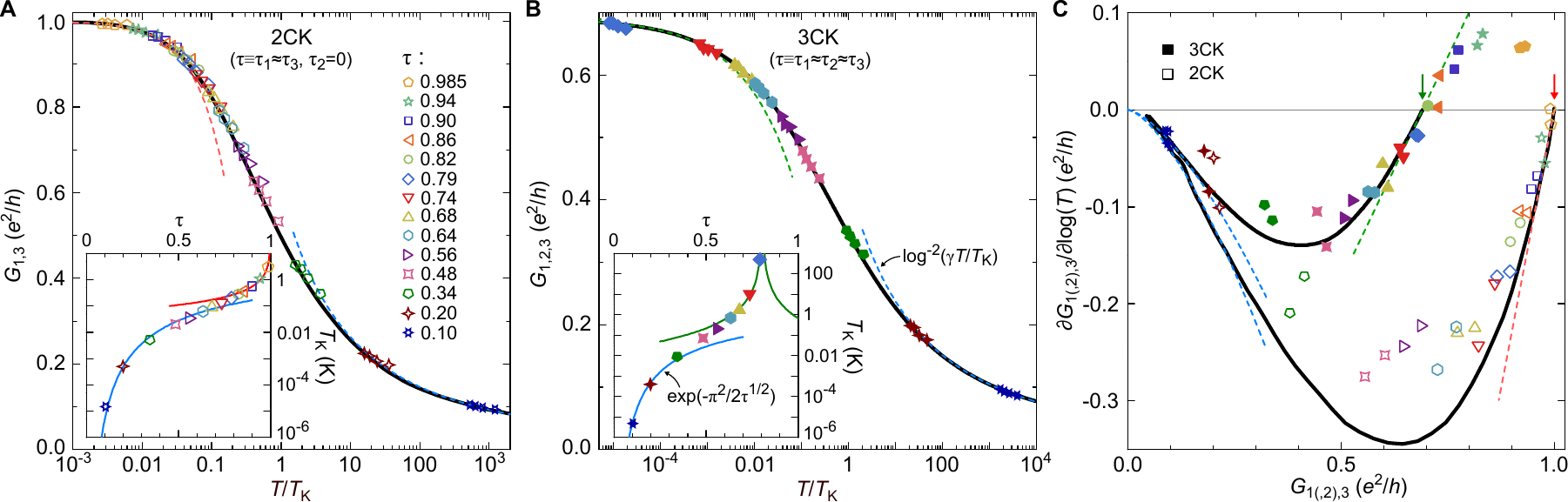}
\caption{
\footnotesize
\textbf{Universal renormalization flow to quantum criticality}.
The measured conductance of the two (A,C) or three (B,C) connected, symmetric channels is shown as symbols (open/full for 2CK/3CK) for a broad range of settings $\tau$.
Black continuous lines are NRG calculations of the universal renormalization flows (2CK/3CK in (A,C)/(B,C)).
Colorized dashed lines shown at low $T/T_\mathrm{K}$ (A,B), and close to the Kondo fixed points (C), display the predicted low-temperature power-laws for 2CK (red in (A,C)) and 3CK (green in (B,C)). 
Light-blue dashed lines shown at large $T/T_\mathrm{K}$ (A,B), and for small channels conductance (C), represent the predicted high-temperature logarithmic scaling $G_{1(,2),3}\propto\log^{-2}(\gamma T/T_\mathrm{K}\gg1)$, with the slightly different 2CK and 3CK prefactors and $\gamma$ here used as fit parameters.
\textbf{A}, \textbf{B}, Data ($T\in\{7.9,9.5,12,18\}$\,mK) and predictions are plotted versus $T/T_\mathrm{K}$ in $\log$ scale.
The corresponding experimental $T_\mathrm{K}$ are shown in insets as symbols versus $\tau$, together with theoretical predictions for tunnel contacts $\tau\ll1$ (light-blue continuous lines) and for very large $T_\mathrm{K}$ at $|\tau-\tau_\mathrm{c}|\ll1$ (red/green continuous line for 2CK/3CK in insets of (A)/(B)).
\textbf{C}, Direct data-theory comparison (no $T/T_\mathrm{K}$ rescaling) with $\partial G_{1(,2),3}/\partial\log(T)$ plotted versus $G_{1(,2),3}$.
The discrete experimental differentiation is performed with measurements at $T\in\{7.9,12,18\}$\,mK.
Kondo fixed points are indicated by arrows.
}
\label{Fig_NRG}
\end{figure*}

{\noindent\textbf{Universal scalings toward quantum criticality.}}\\
First, the power-law exponents when approaching the 2CK and 3CK low-temperature fixed points are characterized, and found different from the characteristic $T^2$ for Fermi liquids.
For this purpose, the distance $\Delta G$ between measured $G_{1(,2),3}$ and the theoretically predicted fixed point $G_\mathrm{2CK}$ ($G_\mathrm{3CK}$) is plotted in Fig.~3 versus $T/T_\mathrm{K}$.
The continuous straight lines show the universal power-law scalings asymptotically predicted at low $T/T_\mathrm{K}$ for the conductance in the present `charge' Kondo implementation: $\Delta G\propto T$ for 2CK \cite{Furusaki1995b,Mitchell2016,Landau2017}, $\Delta G\propto T^{2/5}$ for 3CK \cite{Cox1998,Affleck1993,supplementary} (see \cite{supplementary} for further discussion).
Comparing with the data requires to fix for each $\tau$ the corresponding scaling Kondo temperature $T_\mathrm{K}(\tau)$.
Symbols in the insets represent the experimentally extracted values of $T_\mathrm{K}$ versus $\tau$, which were obtained in practice by matching the lowest temperature data point for each tuning of $\tau$ with the displayed theoretical power-law.
The data-theory comparison in the main panel is therefore in the conductance evolution as temperature is increased.
We find that the experiment is consistent with predictions close enough to the fixed points ($\Delta G\lesssim0.1e^2/h$).
Note that the precision is here limited by the increasing relative experimental uncertainty as $\Delta G$ is reduced.
A direct extraction of the temperature exponents from the $\Delta G<0.1e^2/h$ data at $T\in\{7.9,12\}$\,mK (satisfying the NRG universality criteria $T\lesssim E_\mathrm{C}/20k_\mathrm{B}\simeq15$\,mK) gives $\alpha_\mathrm{2CK}=0.83\pm0.08$ for 2CK and $\alpha_\mathrm{3CK}=0.42\pm0.17$ for 3CK.

Second, the investigation is extended to the full 2CK and 3CK universal renormalization flows.
Measurements (symbols) are now confronted in Fig.~4A,B,C with NRG calculations spanning the whole range of $T/T_\mathrm{K}$ (continuous black lines, see \cite{supplementary}).
In panel A (B), $G_{1(,2),3}$ is plotted versus $\log(T/T_\mathrm{K})$.
Following standard procedures, the theoretical scaling Kondo temperature $T_\mathrm{K}$ was normalized so that the NRG universal conductance takes a value equal to half that of the Kondo fixed point at $T=T_\mathrm{K}$.
As in Fig.~3, the experimental $T_\mathrm{K}(\tau)$ (symbols in inset) are adjusted by matching data with theory at $T\simeq7.9$\,mK.
These $T_\mathrm{K}(\tau)$ remain therefore identical to those in the insets of Fig.~3 as long as NRG calculations and asymptotic power-laws are indistinguishable (i.e. for $T_\mathrm{K}\gg7.9$\,mK).
Remarkably, we observe a quantitative agreement data-universal NRG prediction over six (2CK) or eight (3CK) orders of magnitude in $T/T_\mathrm{K}$.
Figure~4C shows a direct comparison of the same measurements and predictions in a scale-invariant representation, that does not involve rescaling the temperature in units of $T_\mathrm{K}$, by displaying $\partial G_{1(,2),3}/\partial\log(T)$ versus $G_{1(,2),3}$.
In this representation, data points correspond to experimental measurements of the so-called beta-function that determines the corresponding 2CK or 3CK renormalization group equation for the conductance.
The straight dashed lines near 2CK and 3CK fixed points (arrows) represent the predicted non-Fermi liquid power-law behaviors discussed in the previous paragraph.
Comparing with the experimental slope therefore complement the approach in Fig.~3. 
Note the experimental `analogue quantum simulation' of the universal 3CK beta-function at $G_{1,2,3}>G_\mathrm{3CK}$, out of reach of NRG calculations.

Third, we explore and understand the quantitative relationship between scaling Kondo temperature $T_\mathrm{K}$ and microscopic model parameter $\tau$ (insets of Figs.~3 and 4).
At small $\tau\lesssim0.5$, the same expected exponential behavior $T_\mathrm{K}\simeq (E_\mathrm{C}/10k_\mathrm{B})\exp(-\pi^2/\sqrt{4\tau})$ is observed for 2CK and 3CK \cite{Furusaki1995b}.
At larger $\tau$, $T_\mathrm{K}$ appears to diverge at a specific setting $\tau_\mathrm{c}$.
As Kondo physics emerges only for $T<E_\mathrm{C}/k_\mathrm{B}\simeq300$\,mK, a much larger value of the extracted scaling Kondo temperature $T_\mathrm{K}$ implies that only the low-temperature part ($T/T_\mathrm{K}\ll1$) of the full universal scaling curve is accessible with the corresponding device setting \cite{supplementary}.
For 2CK, theory predicts a divergence at $\tau_\mathrm{c}=1$ as $T_\mathrm{K}(1-\tau\ll1)\propto1/(1-\tau)$, which is displayed by the identical continuous red lines in the insets of Figs.~3 and 4A \cite{Furusaki1995b} (see also the prediction of a peaked $T_\mathrm{K}(J)$ in \cite{Lebanon2003,Kolf2007}).
For 3CK, the observed value $\tau_\mathrm{c}\simeq0.8$ is higher than $G_\mathrm{3CK}h/e^2\simeq0.69$.
This is due to the conductance suppression by Coulomb interaction at temperatures $T\gtrsim E_\mathrm{C}/k_\mathrm{B}$, prior to the development of universal Kondo physics at low temperatures.
Assuming theoretically that $T_\mathrm{K}$ diverges at $\tau_\mathrm{c}$, we generally find \cite{supplementary} that a low-temperature conductance power law $\Delta G\propto T^\alpha$ corresponds to a power law divergence as $T_\mathrm{K}\propto|\tau-\tau_\mathrm{c}|^{-1/\alpha}$.
The observed close agreement between experimental $T_\mathrm{K}(|\tau-\tau_\mathrm{c}|\ll1)$ in 2CK and 3CK configurations with, respectively, $T_\mathrm{K}\propto|\tau-1|^{-1}$ (red lines in insets) and $T_\mathrm{K}\propto|\tau-0.8|^{-5/2}$ (green lines in insets), therefore further establishes the predicted non-Fermi liquid Kondo exponents for two ($\alpha_\mathrm{2CK}=1$) and three ($\alpha_\mathrm{3CK}=2/5$) symmetric channels.\\

\begin{figure*}[!thb]
\renewcommand{\figurename}{Fig.}
\centering
\includegraphics[width=120mm]{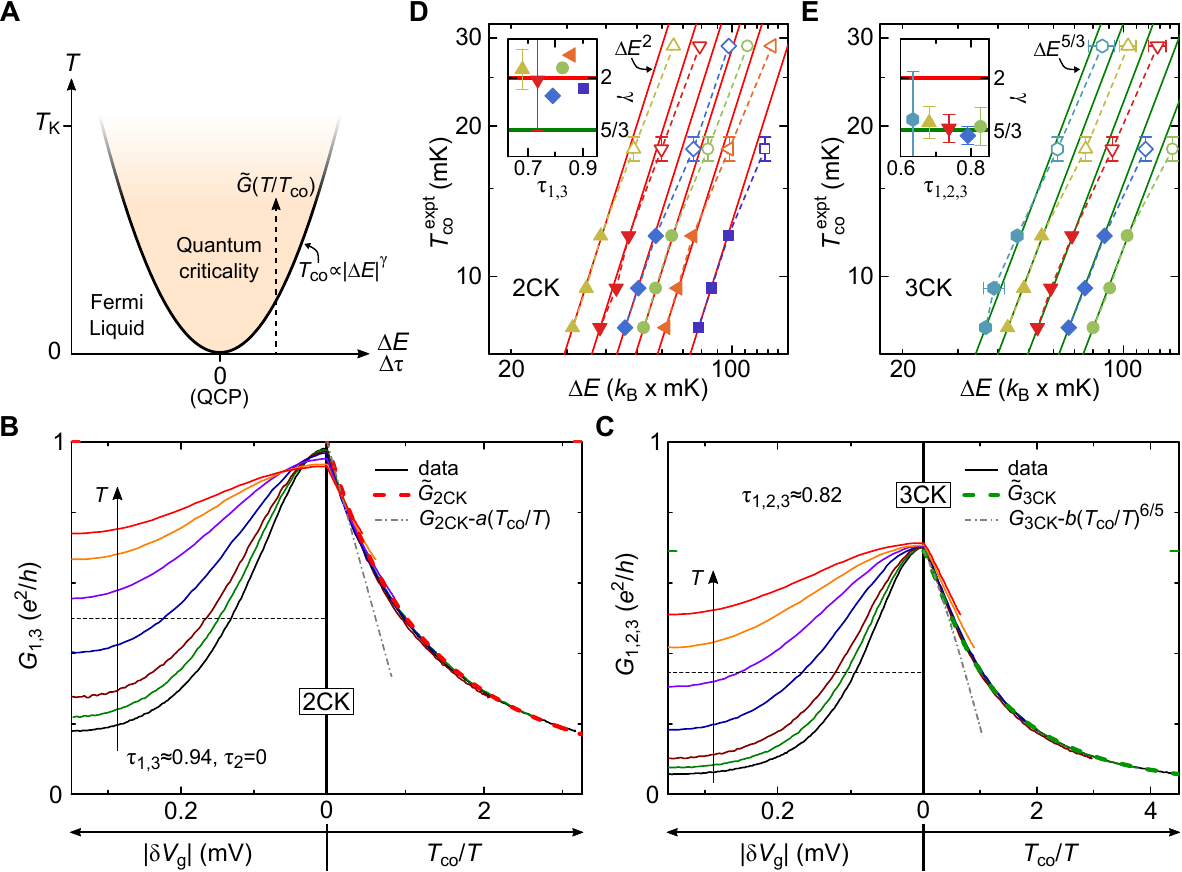}
\caption{
\footnotesize
\textbf{Crossover from quantum criticality by pseudospin degeneracy breaking}. 
\textbf{A}, Quantum criticality extends as $T$ rises.
It is delimited from below by the crossover temperature $T_\mathrm{co}$, which increases as a power-law for small parameter-space distances from the critical point (e.g. charge pseudospin energy splitting $\Delta E\propto\delta V_\mathrm{g}$, channels asymmetry $\Delta\tau$).
Along the crossover, theory predicts universal $T/T_\mathrm{co}$ scalings (e.g. $G_i(T,\Delta E)=\tilde{G}(T/T_\mathrm{co})$).
\textbf{B}, \textbf{C}, The conductance of two (B) and three (C) symmetric channels set, respectively, to $\tau_{1,3}\simeq0.94$ and $\tau_{1,2,3}=0.82$, are plotted as continuous lines versus $|\delta V_\mathrm{g}|$ (left side) and $T_\mathrm{co}/T$ (right side, see text) for $T\in\{7.9,9.5,12,18,29,40,55\}$\,mK.
Colored thick dashed lines (grey dash-dotted lines) shown in right sides display the corresponding theoretical universal crossover curve $\tilde{G}_\mathrm{2CK}$ and $\tilde{G}_\mathrm{3CK}$ (the predicted $T_\mathrm{co}/T\ll1$ power-laws).
The only fit parameter is an unknown fixed prefactor for the 3CK crossover scale $T_\mathrm{co}$ (no fit parameters in (B), see text).
\textbf{D}, \textbf{E}, Experimental crossover temperatures $T_\mathrm{co}^\mathrm{expt}$ are plotted as symbols in a $\log$-$\log$ scale versus $\Delta E$, for two (D) and three (E) symmetric channels.
Each set of symbols connected by dashed lines represents one device setting $\tau_{1(,2),3}$ (see insets).
Full symbols correspond to $T_\mathrm{co}^\mathrm{expt}\leq12$\,mK.
Straight continuous lines display the predicted power-laws $T_\mathrm{co}\propto\Delta E^\gamma$, with $\gamma=2(5/3)$ for 2CK (3CK).
Fitting $T_\mathrm{co}^\mathrm{expt}(\Delta E)\leq12$\,mK separately for each $\tau$ yields the values of $\gamma$ shown as symbols in the insets with the fit standard error.
}
\label{Fig_co_dE}
\end{figure*}

{\noindent\textbf{Crossover from quantum criticality.}}\\
When does quantum criticality apply? 
As temperature is increased (up to some limit, here $T\lesssim\min(T_\mathrm{K},E_\mathrm{C}/k_\mathrm{B})$), quantum criticality is generally expected to span over a larger range of system parameters, away from the $T=0$ quantum critical point (Fig.~5A).
The so-called crossover temperature $T_\mathrm{co}$ delimits quantum criticality from below, with the critical point itself corresponding to $T_\mathrm{co}=0$.
Generically, the crossover from quantum criticality as temperature is lowered should follow universal curves versus the reduced parameter $T/T_\mathrm{co}$.
Indeed, $T_\mathrm{co}$ is the only relevant temperature scale, encapsulating all microscopic details, provided the high-energy cutoff for quantum criticality is much higher.
In tunable circuits, the crossover from 2CK quantum criticality was explored versus Kondo channels asymmetry \cite{Keller2015,Iftikhar2015} and, in the different implementation of a spin-polarized quantum dot embedded into a dissipative circuit, versus the difference between resonant dot level and Fermi energy \cite{Mebrahtu2013}.
These experiments corroborate the existence of a universal $T/T_\mathrm{co}$ scaling, as well as the predicted quadratic increase of $T_\mathrm{co}$ for small deviations from the 2CK critical point \cite{Cox1998,Pustilnik2004b,Sela2011}.
Here we explore the disparate universal and exotic behaviors along the different crossovers induced by breaking the Kondo (pseudo)spin degeneracy or the channel symmetry, observe the development of the quantum phase transition across the symmetric 3CK quantum critical point, and demonstrate `super-ballistic' conductances.

In a first step, we investigate the crossover from 2CK and 3CK quantum criticality induced by breaking the energy degeneracy of the Kondo impurity, with the connected channels remaining symmetric.
We establish (\textit{i}) the different 2CK and 3CK power-law dependence $T_\mathrm{co}\propto|\Delta E|^\gamma$ for small energy splitting of the charge pseudospin $\Delta E=2 E_\mathrm{C}\delta V_\mathrm{g}/\Delta\ll E_\mathrm{C}$ with $\Delta\simeq0.7$\,mV the gate voltage period; (\textit{ii}) a generalized expression of $T_\mathrm{co}$ for arbitrary $\Delta E$; (\textit{iii}) the theoretical universal crossover curves $\tilde{G}_\mathrm{2CK}(T/T_\mathrm{co})$ and $\tilde{G}_\mathrm{3CK}(T/T_\mathrm{co})$, obtained analytically in \cite{Furusaki1995b,Mitchell2016} for 2CK and by NRG here for 3CK.

The crossover temperature $T_\mathrm{co}$ is defined such that the conductance is halfway between the quantum critical regime ($\approx G_\mathrm{2CK(3CK)}$, at $T_\mathrm{co}\ll T\ll T_\mathrm{K}$) and the Fermi liquid regime ($\approx0$, at $T_\mathrm{co}\gg T$), i.e. $G_{1(,2),3}(\Delta E,T=T_\mathrm{co})\equiv G_\mathrm{2CK(3CK)}/2$.
In practice, we fix the electronic temperature $T$ and adjust the energy splitting $\Delta E\propto\delta V_\mathrm{g}$ in order to obtain this midway conductance value, if possible.
In Fig.~5B,C, this corresponds to the crossings between continuous and horizontal dashed lines, where the experimentally extracted crossover temperature directly reads $T_\mathrm{co}^\mathrm{expt}(\Delta E)=T$.
Symbols in Fig.~5D,E display $T_\mathrm{co}^\mathrm{expt}$ versus $\Delta E$ for the settings $\tau$ where $T_\mathrm{co}\propto\Delta E^\gamma$ is expected \cite{supplementary}.

The predicted corresponding power-laws are shown as continuous lines ($T_\mathrm{co}\propto\Delta E^2$ for 2CK, $T_\mathrm{co}\propto\Delta E^{5/3}$ for 3CK \cite{Cox1998}).
Fitting separately, for each $\tau$, the $T_\mathrm{co}^\mathrm{expt}(\Delta E)\leq12$\,mK data (fulfilling the universality NRG criteria) yield the values of $\gamma$ displayed as symbols in the insets.
A statistical analysis of these values give $\gamma_\mathrm{2CK}=2.01\pm0.04$ and $\gamma_\mathrm{3CK}=1.69\pm0.02$ for the crossovers from 2CK and 3CK, respectively, in close agreement with theory.

The theoretically predicted universal crossover curves $\tilde{G}_\mathrm{2CK}(T_\mathrm{co}/T)$ and $\tilde{G}_\mathrm{3CK}(T_\mathrm{co}/T)$, shown as thick dashed lines in the right panels of Fig.~5B,C, are confronted with conductance data.
Continuous lines in the left panels represent the conductance measured at different temperatures versus gate voltage for $\tau_{1,3}\simeq0.94$ (Fig.~5B) and $\tau_{1,2,3}\simeq0.82$ (Fig.~5C).
These settings correspond to well-developed quantum critical regimes $T\ll T_\mathrm{K}$ (small $\Delta G$), a necessary condition to investigate $\tilde{G}_\mathrm{2CK}$ and $\tilde{G}_\mathrm{3CK}$ down to small $T_\mathrm{co}/T$.
As shown in the right panels, the gate voltage sweeps at different temperatures (continuous lines) are superimposed when plotted versus calculated $T_\mathrm{co}/T$, thereby demonstrating the predicted universal character of the crossover from quantum criticality.
Moreover, we find a precise match between experimental universal curves and theoretical predictions $\tilde{G}_\mathrm{2CK}$ and $\tilde{G}_\mathrm{3CK}$.
Note that $T_\mathrm{co}$ is obtained from experimental parameters using generalized expressions that remain valid for arbitrary gate voltage, beyond the power-law at small detuning.
For `charge' 2CK device with near ballistic channels, the full quantitative expression derived in \cite{Furusaki1995b} was used in Fig.~5B: $T_\mathrm{co}\simeq1.444E_\mathrm{C}(1-\tau_{1,3})\sin^2(\pi\delta V_\mathrm{g}/\Delta)$.
The data-$\tilde{G}_\mathrm{2CK}$ comparison in Fig.~5B is therefore without any fit parameter.
For 3CK, we expect from NRG calculations the similar generalization $T_\mathrm{co}=\lambda_\mathrm{3CK}\sin^{5/3}(\pi\delta V_\mathrm{g}/\Delta)$ \cite{supplementary}, which was used Fig.~5C.
As the prefactor $\lambda_\mathrm{3CK}(\tau,E_\mathrm{C})$ is not known, the value $\lambda_\mathrm{3CK}=36$\,mK was freely adjusted in the data-$\tilde{G}_\mathrm{3CK}$ comparison shown in Fig.~5C.

\begin{figure}[!thb]
\renewcommand{\figurename}{Fig.}
\centering
\includegraphics[width=89mm]{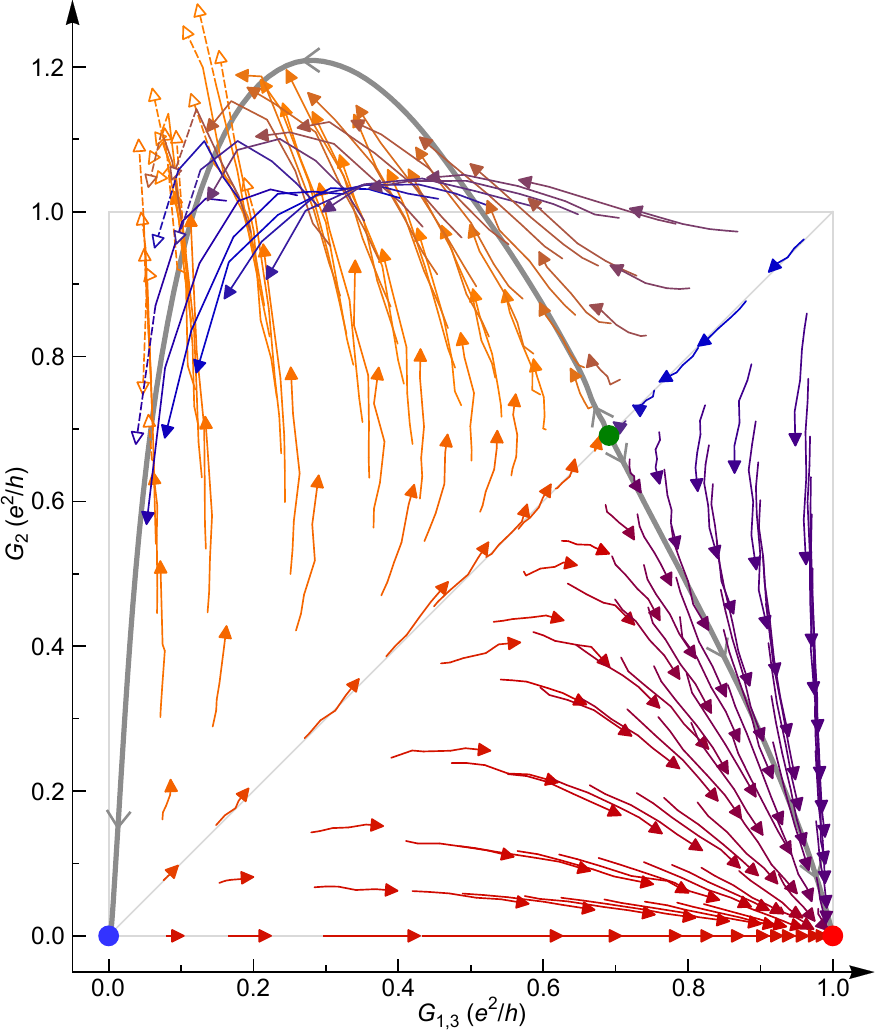}
\caption{
\footnotesize
\textbf{Three-channel Kondo renormalization flow with super-ballistic conductances}. 
Each colored line displays for a fixed device tuning ($\{\tau_1\simeq\tau_3,\tau_2\}$) at charge degeneracy ($\delta V_\mathrm{g}=0$), the measured channels' conductance at $T=55,$ $40,$ $29,$ $18,$ $12,$ and $7.9$\,mK (arrow at lowest $T$).
The lines' color reflects the direction (the angle) of the vector connecting lowest and highest temperature data points, to improve readability.
QPC$_1$ and QPC$_3$ are set symmetric ($\tau_1\approx\tau_3$ tuned among fourteen values from $0.1$ to $0.985$ \cite{supplementary}), and only the renormalized average $G_{1,3}$ is shown on the horizontal axis.
QPC$_2$ is adjusted separately to a coupling $\tau_2$ selected among the same fourteen values and also $\tau_2=0$.
Solid lines and filled arrows indicate an experimental standard error on $G_2h/e^2$ and $G_{1,3}h/e^2$ below $0.05$ (usually well below).
Dashed lines and open arrows indicate a standard error on $G_2h/e^2$ between $0.05$ and $0.1$.
The green, red and blue disks correspond, respectively, to the predicted 3CK, 2CK and 1CK low temperature fixed points.
The thick grey lines represent NRG calculations of the universal crossover flows from 3CK, with arrows pointing to lower temperatures.
Notably, the conductance $G_2$ can markedly exceed the maximum free electron limit $e^2/h$.
}
\label{Fig_co_dtau}
\end{figure}

In a second step, the development of the 3CK QPT driven by the channels' competition to screen the Kondo spin is plainly observed, through the conductance renormalization flow of asymmetric channels upon lowering temperature (Fig.~6).
Here the Kondo `charge' pseudospin is energy degenerate ($\delta V_\mathrm{g}=0$), QPC$_{1,3}$ are tuned symmetric ($\tau_1\simeq\tau_3$), and $\tau_2$ is adjusted separately.
Fig.~6 displays as colored lines the temperature evolution of the measured conductances $G_{2}$ (vertical axis) and $G_{1,3}$ (horizontal axis) from 55 to $7.9$\,mK (arrow at lowest $T$), with each line corresponding to a different device setting.
In total, $15\times14$ settings of $\{\tau_2,\tau_1\simeq\tau_3\}$ were measured, with $\tau_{1,2,3}$ picked among fourteen fixed values ranging from $0.1$ to $0.985$ \cite{supplementary} and including also $\tau_2=0$.
The data closest to the diagonal grey line correspond to three channels tuned symmetric ($\tau_1\simeq\tau_2\simeq\tau_3$).
Below the diagonal, where $\tau_2<\tau_1\simeq\tau_3$, the data flow toward the predicted 2CK fixed point (red disk, at $G_{1,3}=e^2/h$ and $G_2=0$).
Above the diagonal, where  $\tau_2>\tau_1\simeq\tau_3$ such that a flow toward the 1CK fixed point involving QPC$_2$ is expected (blue disk, at $G_{1,2,3}=0$), we observe a monotoneous decrease of the conductance G$_{1,3}$ across the less strongly coupled QPCs.
In contrast, $G_2$ first rises, markedly oversteps the free-electron quantum limit $e^2/h$ (up to $+25\%$), and then decreases toward the zero conductance 1CK fixed point as $T$ is further reduced.

The non-monotonous behavior of $G_2$ when higher than $G_{1,3}$ might appear counter-intuitive.
Indeed, a flow toward the low-temperature 1CK `strong coupling' fixed point is expected, which corresponds to a renormalized Kondo coupling growing monotonously ($J_2\rightarrow\infty$).
However $J_2$ connects with the tunnel coupling/hopping integral of electrons across QPC$_2$ in the `charge' Kondo mapping, and free-electron theory predicts a non-monotonous dependence of the conductance with the hopping integral (with a maximum for the $J_2$ value that best preserves translational invariance, and $G_2(J_2\rightarrow\infty)=0$).
In contrast, the present measurement of a conductance that exceeds the maximum possible value for non-interacting electrons in the ballistic limit is highly non-trivial and was not anticipated (although reproduced by new NRG calculations, see below).
Notably, such a super-ballistic conductance, also in an intermediate temperature range and of similar amplitude, was coincidentally observed in clean graphene constrictions \cite{KrishnaKumar2017} and explained as a collective viscous flow of the electronic fluid induced by electron-electron collisions \cite{Guo2017}.
We speculate that the electron-electron interactions mediated by the Kondo impurity within the electronic channel across QPC$_2$, expected particularly strong near the turning point where $G_2$ is maximum, might also result in such a viscous electronic fluid behavior.
An interesting specificity of our system is that the super-ballistic magnitude and the temperature range where it takes place can be controlled in-situ, by separately adjusting the channels.

The experimental findings are compared with NRG calculations of the universal crossover flow from 3CK quantum criticality, induced by an initially minute asymmetry between $G_2$ and $G_{1,3}$ \cite{supplementary}.
These are displayed as two thick grey lines originating from the 3CK fixed point, with arrows pointing toward lower temperatures.
For $G_{1,3}>G_2$, NRG predicts a monotonous crossover flow from 3CK to 2CK conductance fixed points that closely matches the nearby data.
For $G_2>G_{1,3}$, the universal NRG crossover flow from 3CK to 1CK reproduces the observed non-monotonous behavior, confirms the naively expected vanishing of $G_2$ at the 1CK fixed point, and establishes that a super-ballistic conductance exceeding by approximately $20\%$ the free-electron maximum limit follows from the 3CK model, in quantitative agreement with the experiment.
Note that while experimental and NRG flows point to the same direction near 3CK and 1CK fixed points, clear crossings are also visible in intermediate regimes above the diagonal, including between different experimental device settings.
These mostly take place between flows involving opposite renormalization directions of $G_2$, as expected from the non-monotonous relationship between $G_2$ and Kondo coupling $J_2$ that specifically shows up above the diagonal.\\

{\noindent\textbf{Outlook.}}\\
Quantum impurity models such as the preeminent Kondo-type allow exploring strongly-correlated and quantum critical physics \cite{Vojta2006,Bulla2008}, and their solutions can be used to obtain the properties of strongly-correlated materials within  the `dynamical mean field' approximation \cite{Kotliar2004}.
We have shown that metal-semiconductor hybrid circuits constitute widely tunable and fully characterized `charge' Kondo devices, thereby providing experimental test-beds of exotic many-body behaviors at the high-precision level of quantum simulators.
In particular, the specific observation of super-ballistic conductance opens a research path for low-power electronics.
Although the present implementation has no clear application potential, it forms a powerful platform to understand the underlying mechanisms of a behavior susceptible to arise in diverse clean systems with strong electron-electron interactions.
We anticipate that similar metal-semiconductor hybrids will form building blocks for a wide range of investigations of the strongly-correlated electron physics, and in particular the emergence of exotic parafermions quasi-particles \cite{Emery1992,Affleck2001,Landau2017}.
Measurements of complementary observables such as charge susceptibility, fluctuations and heat current, as well as investigations of the dynamical and out-of-equilibrium responses could unveil yet hidden facets of the exotic underlying physics.
Furthermore, direct generalizations of the present `charge' Kondo implementation should grant access to quantitative investigations of many thus far inaccessible strongly-correlated phenomena \cite{Vojta2006}, including the nano-engineered competition between Kondo channels, dissipation (note the specific proposal in \cite{LeHur2004}), fractional quantum Hall effect and multiple impurities.


\subsection*{Acknowledgments}
We thank I.~Affleck, S.~Florens, L. Fritz, L.~Glazman, C.~Kane, K.~Matveev, Y.~Nazarov, E.~Sela, G.~Zarand, F.~Zhang for discussions.
This work was supported by the French RENATECH network and the national French program `Investissements d'Avenir' (Labex NanoSaclay, ANR-10-LABX-0035).\\
Authors contributions: Z.I. and F.P. performed the experiment; Z.I., A.A. and F.P. analyzed the data; A.K.M. performed the NRG developments; C.M and P.S extended the analytical predictions; F.D.P. fabricated the sample with inputs from A.A.; U.G., A.C. and A.O. grew the 2DEG; F.P. led the project and wrote the manuscript with inputs from Z.I., A.A., A.K.M, U.G, C.M. and P.S.\\
Correspondence and requests for materials should be addressed to F.P. (frederic.pierre@u-psud.fr).

\end{document}


\title{Supplementary Materials for\\Tunable Quantum Criticality and Super-ballistic Transport in a `Charge' Kondo Circuit}

\author{Z. Iftikhar}
\affiliation{Centre de Nanosciences et de Nanotechnologies (C2N), CNRS, Univ Paris Sud, Universit\'e Paris-Saclay, 91120 Palaiseau, France}
\author{A. Anthore}
\affiliation{Centre de Nanosciences et de Nanotechnologies (C2N), CNRS, Univ Paris Sud, Universit\'e Paris-Saclay, 91120 Palaiseau, France}
\affiliation{Univ Paris Diderot, Sorbonne Paris Cit\'e, 75013 Paris, France}
\author{A.K. Mitchell}
\affiliation{School of Physics, University College Dublin, Dublin 4, Ireland}
\author{F.D. Parmentier}
\affiliation{Centre de Nanosciences et de Nanotechnologies (C2N), CNRS, Univ Paris Sud, Universit\'e Paris-Saclay, 91120 Palaiseau, France}
\author{U.~Gennser}
\affiliation{Centre de Nanosciences et de Nanotechnologies (C2N), CNRS, Univ Paris Sud, Universit\'e Paris-Saclay, 91120 Palaiseau, France}
\author{A. Ouerghi}
\affiliation{Centre de Nanosciences et de Nanotechnologies (C2N), CNRS, Univ Paris Sud, Universit\'e Paris-Saclay, 91120 Palaiseau, France}
\author{A. Cavanna}
\affiliation{Centre de Nanosciences et de Nanotechnologies (C2N), CNRS, Univ Paris Sud, Universit\'e Paris-Saclay, 91120 Palaiseau, France}
\author{C. Mora}
\affiliation{Laboratoire Pierre Aigrain, Ecole Normale Sup\'{e}rieure-PSL Research University, CNRS, UPMC Univ Paris 06-Sorbonne Universit\'es, Univ Paris Diderot-Sorbonne Paris Cit\'e, 75005 Paris, France}
\author{P. Simon}
\affiliation{Laboratoire de Physique des Solides, CNRS, Univ. Paris-Sud, Universit\'e Paris-Saclay, 91405 Orsay, France}
\author{F. Pierre}
\affiliation{Centre de Nanosciences et de Nanotechnologies (C2N), CNRS, Univ Paris Sud, Universit\'e Paris-Saclay, 91120 Palaiseau, France}

\maketitle

\section*{Materials and Methods}

\subsection{Sample and setup}

\begin{figure}[htb]
\renewcommand{\thefigure}{S\arabic{figure}}
\renewcommand{\figurename}{Fig.}
\centering\includegraphics [width=\columnwidth]{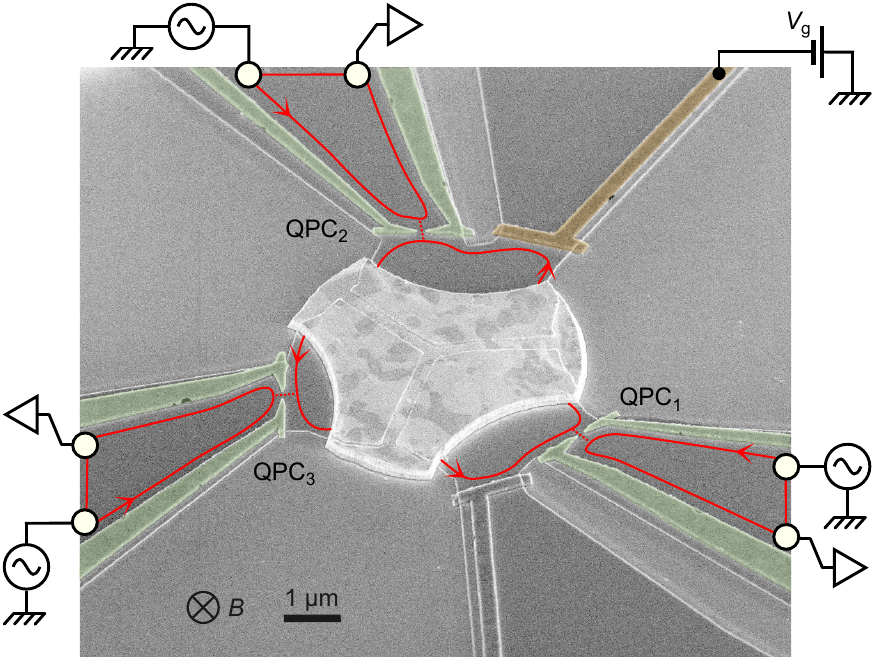}
\caption{\small
\textbf{Colored micrograph of the sample}, with measurement schematic. 
The central metallic island (bright) is connected to the circuit through QPCs formed by field effect in a buried two-dimensional electron gas (dark grey) using surface split gates (green).
The voltage $V_\mathrm{g}$, used to tune the metallic island at charge degeneracy, is applied to a capacitively coupled plunger gate (yellow).
Note that the voltages applied to the two lateral gates (yellow and uncolored) are sufficiently negative to deplete the two-dimensional electron gas underneath (except for electronic thermometry and characterization purposes).
Due to the strong perpendicular magnetic field $B=2.7\,$T, the current propagates along spin-polarized edge channels (red lines) of the integer quantum Hall effect (only the relevant outermost edge channel is shown).
}
\end{figure}

{\noindent\textbf{Sample.}} 
The sample was nanostructured by standard e-beam lithography in a Ga(Al)As two-dimensional electron gas located 105\,nm below the surface, of density $2.5\,10^{11}\,\mathrm{cm}^{-2}$ and mobility $10^6\,\mathrm{cm}^2\mathrm{V}^{-1}\mathrm{s}^{-1}$. 
The ohmic contact between the micrometer-scale metallic node/`island' and the buried two-dimensional electron gas was realized by thermal diffusion into the semiconductor of a metallic multilayer of nickel (30\,nm), gold (120\,nm) and germanium (60\,nm).
The typical energy spacing between electronic levels in the central metallic island is $\delta\approx k_\mathrm{B}\times0.2\,\mu$K.
The measurements were performed under a strong magnetic field $B\approx2.7\,$T, in the regime of the integer quantum Hall effect at filling factor $\nu=3$.
This specific choice was driven by the need to have simultaneously three well-behaved QPCs, without sharp resonances over the full range $\tau\in[0,1]$, as well as very low temperatures (vibrations in our setup increase the temperature at high field, see [33]) and good ohmic contacts between the outer edge channel and the metallic island.
The interface quality between metallic island and two-dimensional electron gas is characterized by a residual reflection probability lower than $4\,10^{-4}$ for each of the outermost quantum Hall edge channels originating from the three QPCs.
The charging energy $E_\mathrm{C}=e^2/2C\approx k_\mathrm{B}\times299\pm5\,$mK $\approx25.8\pm0.5\,\mu$eV is obtained from the half-height in drain-source dc bias voltage of measured Coulomb diamonds (not shown, the uncertainties are rough estimates based on four different measurements).\\

{\noindent\textbf{Experimental setup.}} 
The device was installed in a dilution refrigerator including multiple filters along the electrical lines and two shields at the mixing chamber.
Details on the fridge wiring and on the sample holder are provided in [33].
Conductance measurements were carried out by standard lock-in techniques at low frequencies, below 200\,Hz (see sample micrograph with a schematic measurement setup in Fig.~S1).
The amplification gains and injected signals are precisely calibrated on-chip, from the signal measured with the QPCs closed ($\tau=0$).
Noise measurements for the electronic thermometry were performed in the MHz range using a homemade cryogenic amplifier.
Details on a very similar noise measurement setup can be found in the supplementary materials of \cite{Jezouin2013b}.\\

\subsection{Experimental methods}

{\noindent\textbf{Electronic temperature.}} 
Below 50\,mK, the electronic temperature was extracted on-chip using quantum shot-noise primary thermometry [33].
Above 50\,mK, $T$ is given by a standard (RuO$_2$) thermometer thermally anchored to the mixing chamber of the dilution refrigerator.
From shot-noise thermometry, we obtain for the data points used in the paper $T=\{7.9\pm0.1, 9.5\pm0.2, 12.1\pm0.2, 18\pm1, 28.9\pm0.8, 40.1\pm0.4\}$\,mK with the uncertainty corresponding to the standard error on the mean value of $T$ determined from the statistical analysis of several temperature measurements (typically 10).
From RuO$_2$ thermometry, we obtain $T=\{40.3\pm0.2, 55.1\pm0.3, 75.4\pm0.6\}$\,mK with the displayed uncertainty corresponding to the temperature drift during the measurements.
Note the good agreement between the two thermometry methods at $40$\,mK.\\

\begin{figure*}[htbp]
\renewcommand{\thefigure}{S\arabic{figure}}
\renewcommand{\figurename}{Fig.}
\centering\includegraphics [width=0.95\textwidth]{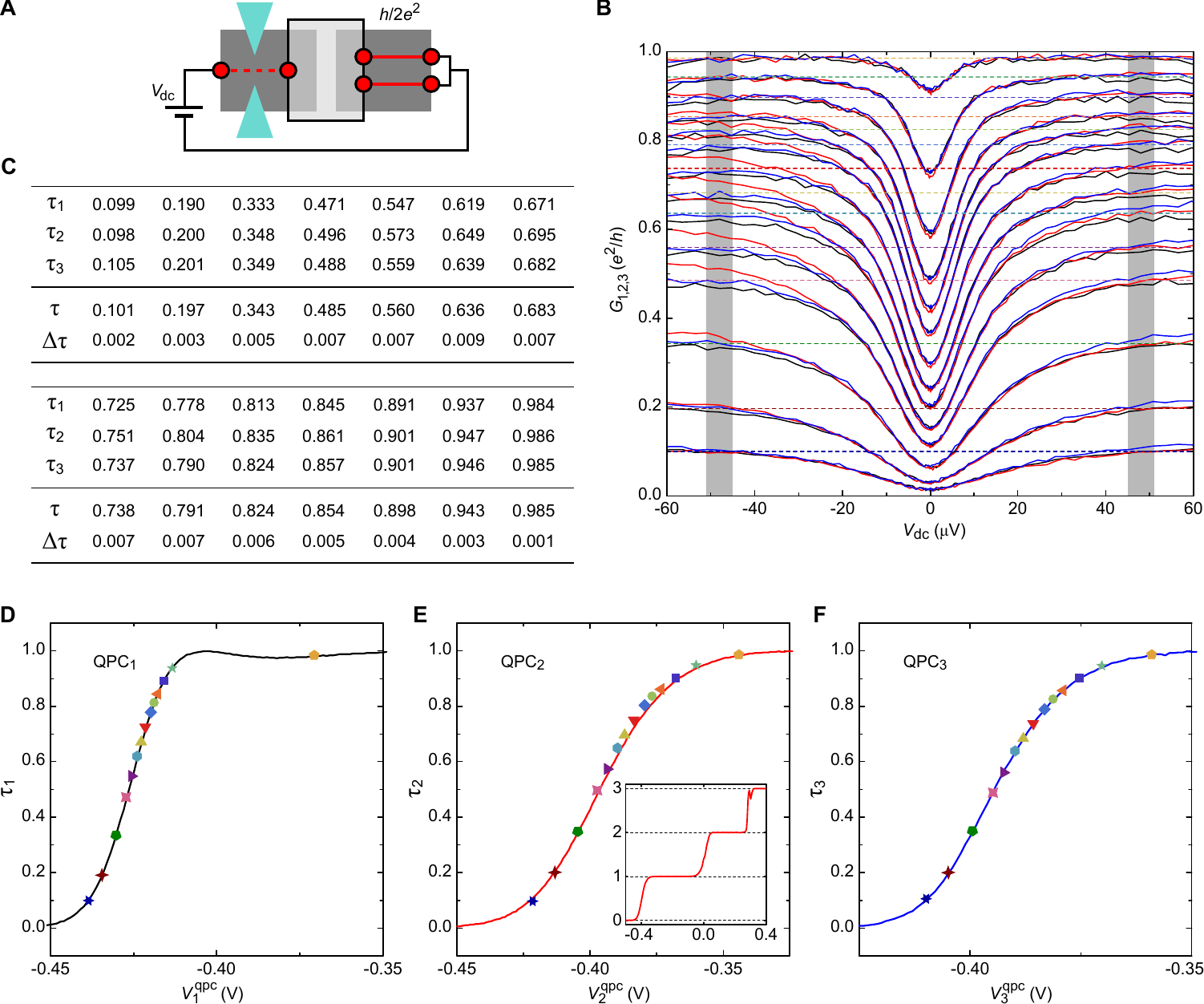}
\caption{\small
\textbf{QPCs characterization.}
\textbf{A}, Schematic circuit used to determine $\tau_{1,2,3}$: a large bias voltage is applied to the characterized QPC in series with a known resistance $h/2e^2$. 
\textbf{B}, Conductance of the QPCs measured at $T=18$\,mK versus dc voltage (continuous lines, black for QPC$_1$, red for QPC$_2$ and blue for QPC$_3$), in the configuration shown in (A) with the series resistance subtracted.
The low bias conductance dips result from the dynamical Coulomb blockade, while the high bias plateaus correspond to the `intrinsic' transmission probabilities $\tau_{1,2,3}$.
\textbf{C}, The `intrinsic' transmission probabilities $\tau_{1,2,3}$ at the experimental set points used in the main text are defined as the average of the QPCs conductance on the large bias ranges $\pm [45\,\mathrm{\mu V}, 51\,\mathrm{\mu V}]$ (grey areas in (B)). 
The individual transmission probabilities $\tau_{1,2,3}$ are averaged to give $\tau$ (horizontal colored dashed lines in (B)) and the estimated uncertainty $\Delta\tau$ (the standard error on the mean value $\tau$ calculated from six measurements, at negative and positive bias voltage for the three QPCs).
\textbf{D}, \textbf{E}, \textbf{F}, The `intrinsic' transmissions $\tau_{1,2,3}$ of the QPCs are plotted as symbols versus the voltage $V^\mathrm{qpc}_{1,2,3}$ applied on one side of the corresponding split gate.
The continuous lines are measured using the lateral characterization gates to short-circuit the metallic island for (D),(E), or with $h/2e^2$ in series (see (A)) at an applied bias voltage $V_\mathrm{dc}=-50\,\mu$V for (F).
The inset in (E) shows QPC$_2$ conductance over a larger gate voltage range including several plateaus, which illustrates the broad separation between channels.
}
\end{figure*}

{\noindent\textbf{Quantum point contact characterization.}} 
As generally observed in the integer quantum Hall regime, the conductance across each QPC shows a broad and very well defined $e^2/h$ plateau when varying the corresponding split gate voltage (see inset in Fig.~S2E for a gate voltage sweep of QPC$_2$ including several plateaus).
Tuning the QPC to a conductance below $e^2/h$ corresponds to partially opening a single (spin polarized) electronic channel.
We extract its `intrinsic' transmission probability $\tau$ essentially by applying a large dc bias voltage ($\sim50\,\mu$V) that suppresses the Kondo and Coulomb conductance renormalizations, such that $\tau\simeq Gh/e^2$.

The precise procedure is now described in more details.
For characterizing QPC$_i$ ($i\in\{1,2,3\}$), we close QPC$_{j\neq i}$ ($G_{j\neq i}=0$) and tune a continuous lateral gate on the other side of the metallic island in the middle of a very broad conductance plateau for which two edge channels are perfectly transmitted (the yellow gate in Fig.~S1 for $i\in\{1,3\}$, the uncolorized gate for $i=2$).
Note that the small capacitive crosstalk effect on the characterized QPC is compensated (see next section).
A very well defined and voltage independent (in the probed range) $h/2e^2$ resistor is therefore in series with QPC$_i$, as schematically represented in Fig.~S2A. 
Applying a dc voltage $V_\mathrm{dc}$ across the whole device suppresses the low bias dynamical Coulomb blockade reduction of $G_i$ (see e.g. \cite{SCT1992} for a theoretical description of dynamical Coulomb blockade), as can directly be seen Fig.~S2B.
The `intrinsic' transmission probability $\tau_i$ is here identified with the differential conductance $G_i$ measured at large bias voltage (compared to $E_\mathrm{C}/e$).
The corresponding $G_i(V_\mathrm{dc})$ data at $T=18$\,mK are shown as continuous lines in Fig.~S2B for the different device settings (black, red, blue for $i=1,2,3$, respectively). 
In practice, we extract $\tau_i$ from the average of $G_i(V_\mathrm{dc})$ in the range $|V_\mathrm{dc}|\in[45,51]\,\mu$V (grey bands in Fig.~S2B).
This choice reflects a good compromise between completely suppressing the dynamical Coulomb blockade renormalization (large enough $|V_\mathrm{dc}|$) and making sure that the energy dependence of the intrinsic $\tau_i$ remains negligible (small enough $|V_\mathrm{dc}|$).
Note that the symmetry between QPCs was finely adjusted directly from the conductances measured in the 3CK configuration at the specific temperature $T=18$\,mK.
At large $|V_\mathrm{dc}|\gtrsim20\,\mu$V, visible differences develop both between QPCs as well as between opposite voltages for the same QPC.
We attribute these differences to the small but non-negligible energy dependence of $\tau_i$, which effectively results in an experimental uncertainty on the determination of its absolute value.
The corresponding uncertainty is estimated from the standard error $\Delta\tau$ on the mean value $\tau$ determined from the six measurements $G_{1,2,3}(V_\mathrm{dc}\in\pm[45,51]\,\mu$V) (three QPCs tuned symmetric at low bias, each measured separately for large positive and negative bias voltages).
The mean $\tau$ are shown as horizontal dashed lines in Fig.~S2B.
The extracted values of $\tau_{1,2,3}$, $\tau$ and $\Delta\tau$ are recapitulated in the table shown Fig.~S2C.
Figures~S2D,E,F display $\tau_{1,2,3}$ as symbols versus the voltage $V^\mathrm{qpc}_{1,2,3}$ applied on one side of the split gate controlling QPC$_{1,2,3}$, respectively. 
In order to illustrate the regular gate voltage dependence, continuous traces in the same panels display gate voltage sweeps of the corresponding QPC conductance, with an essentially suppressed dynamical Coulomb blockade renormalization (short-circuiting the island in panels D,E; with a dc voltage of $-50\,\mu$V in panel F).\\

{\noindent\textbf{Capacitive crosstalk corrections.}}
Each QPC is slightly impacted by the gate voltages applied to control the other QPCs.
Thanks to a distance of several microns this capacitive crosstalk is relatively small.
Typically, changing the voltage on the other QPCs' split gates has an effect 100 times smaller than directly changing the voltage of the split gate used to form the considered QPC.
Although small (the effect on $\tau$ is at the most, and generally well below, $0.05$), these capacitive crosstalk corrections were straightforwardly calibrated and systematically corrected for, in order to best preserve the QPCs symmetry.
Typically, spanning $\tau$ on one QPC amounts to gate voltage variations below $0.1$\,V, resulting in crosstalk corrections smaller than 1\,mV on the other QPCs split gate.\\

{\noindent\textbf{Renormalized channel conductance.}}
Here we detail the relation between individual QPC conductances and the currents and voltages across the whole `charge' Kondo device.
Applying a small voltage $V_1$ ($eV_1<k_\mathrm{B}T/2$) on the large voltage biased electrode feeding QPC$_1$ (see Fig.~S1), we measure the current $I_1$ flowing through QPC$_1$ toward the large grounded electrodes on the other sides of QPC$_2$ and QPC$_3$.
From Kirchoff's laws, $I_1/V_1=1/(1/G_1+1/(G_2+G_3))$.
Repeating the same procedure for the three QPCs (in practice this is done simultaneously using lock-in techniques), allows us to extract the individual values of $G_1$, $G_2$ and $G_3$ as long as the three conductances are non-zero, in the 3CK configurations.
Note that in the 2CK configurations ($G_2=0$), the redundant measurements $I_1/V_1=I_3/V_3\equiv G_{1,3}/2$ do not allow us to extract separately $G_1$ and $G_3$.\\

\begin{figure}[htb]
\renewcommand{\thefigure}{S\arabic{figure}}
\renewcommand{\figurename}{Fig.}
\centering\includegraphics [width=\columnwidth]{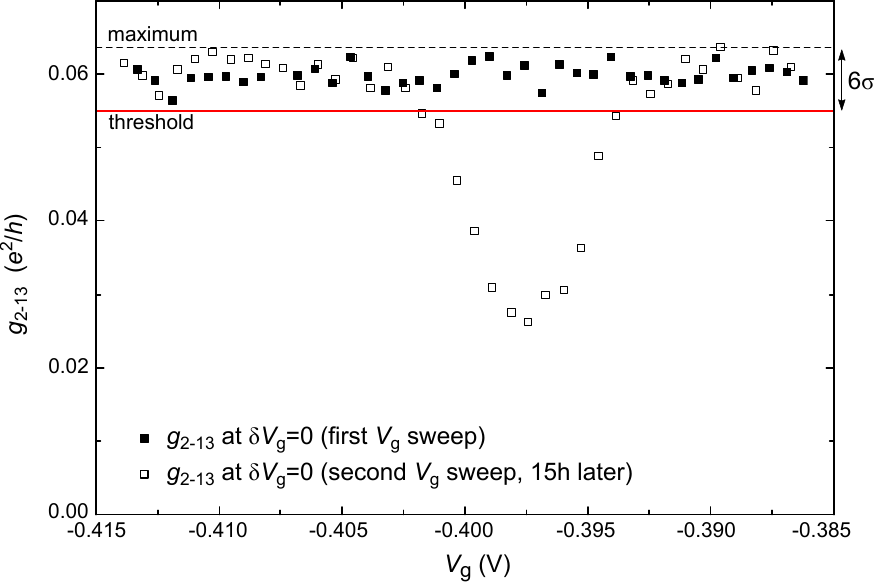}
\caption{\small
\textbf{Data analysis.}
The device conductance through QPC$_2$ ($g_{2-13}\equiv1/(1/G_2+1/(G_1+G_3)$), measured at charge degeneracy ($\delta V_\mathrm{g}=0$) for $\tau_{1,2,3}\approx0.1$ and $T\approx12$\,mK, is plotted as symbols versus gate voltage $V_\mathrm{g}$ at the consecutive peak maximums. 
The same $V_\mathrm{g}$ sweep (a 1h long measurement) is repeated twice with a 15h time interval.
The data points below the statistical threshold shown as a red continuous line are automatically discarded.
\normalsize}
\end{figure}

{\noindent\textbf{Data reproducibility.}}
The experimental data shown in Fig.~2, 3, 4 and 6 are extracted from the average of the conductance at degeneracy ($\delta V_g = 0$) obtained from many Coulomb peaks.
Considering only the symmetric 2CK and 3CK device configurations, a total of 6074 conductance peaks were measured, corresponding in average to 27 peaks per displayed data point.
This allowed for the automatic detection and exclusion of statistically anomalous measurements (of statistical probability $\lesssim1\%$ assuming a gaussian distribution).
The excluded measurement artifacts and the automatic procedure are illustrated in Fig.~S3.
Symbols represent measurements of the maximum peak conductance (at $\delta V_g=0$) across the whole device when the current is injected from the electrode connected to QPC$_2$ ($g_{2-13}\equiv1/(1/G_2+1/(G_1+G_3)$).
The data points obtained by sweeping the gate voltage ($V_g$) at $T\simeq12$\,mK and $\tau\simeq0.1$ are plotted versus $V_g$ at the position of the consecutive peak maximums.
The same sweep was performed twice with a 15h time interval.
While in the first sweep (full symbols) all the maximum peak conductances are at similar values, the second sweep (open symbols) shows a pronounced dip of the measured conductance maximums for $V_g\in[-0.402,-0.393]$\,V.
Such non-reproducible experimental artifacts are attributed to the activation of charge fluctuators in the device vicinity.
In order to discard such dip artifacts, we automatically remove all peak measurements whose conductance is more than six times the expected standard deviation (from instrumental noise, separately calibrated) below the highest measured value.
The highest value and the corresponding lower threshold are shown in Fig.~S3 as a black dashed line and a red continuous line, respectively.\\

{\noindent\textbf{Power-law exponent of crossover temperature.}}
This section concerns the experimental determination of the power-law exponent $\gamma$ describing the increase of the crossover temperature $T_\mathrm{co}$ versus the charge pseudospin energy splitting $\Delta E$ (Fig.~5D,E).
We provide here the explicit device settings criteria used to restrict the analysis where such power-law behavior is expected.
Two additional criteria complement the already mentioned low temperatures corresponding to the universality regime ($T\leq12\text{\,mK}< E_\mathrm{C}/20\simeq 15$\,mK):
(\textit{i}) The power-law dependence only applies in the limit of small energy splitting $\Delta E$, as directly seen from the generalized expression of $T_\mathrm{co}$ for the crossover from 2CK quantum criticality given by Eq.~\ref{TcoFMsym}. 
In practice, we only considered the settings of $\tau$ for which $\Delta E<E_\mathrm{C}/3\simeq k_\mathrm{B}\times100$\,mK (corresponding to a maximum of 9$\%$ relative deviation of $T_\mathrm{co}$ from the power-law dependence).
(\textit{ii}) The Kondo temperature must be sufficiently high with respect to $T$.
Indeed, a universal crossover flow is generally expected only in the limit $T,T_\mathrm{co}\ll T_\mathrm{K}$.
Specifically, a too large difference between $G_{1(,2),3}(\delta V_\mathrm{g}=0)$ and $G_\mathrm{2CK(3CK)}$ (which occurs if $T_\mathrm{K}$ is not large enough compared to $T$) results in a non-negligible gate voltage shift of the crossing point $G_{1(,2),3}(\delta V_\mathrm{g})=G_\mathrm{2CK(3CK)}/2$.
As the gate voltage at the crossing point is used to extract $T_\mathrm{co}(\Delta E)$, such a shift would translate into an experimental error.
In practice, we therefore only considered the settings of $\tau$ for which $|1-G_{1(,2),3}(\delta V_\mathrm{g}=0,T)/G_\mathrm{2CK(3CK)}|<0.25$.\\

\section*{Supplementary text}

\subsection{Model, predictions and super-ballistic observation}

{\noindent\textbf{Multichannel `charge' Kondo model.}}
The mapping of the circuit Hamiltonian in the weak coupling limit to an anistropic Kondo model is here made explicit, following [29].
The circuit Hamiltonian for a metallic island coupled through $N$ single-channel point contacts to $N$ different leads is usually written [29,32]:
\begin{equation}
H=H_I+\sum_{i=1}^{N}\left(H_{Li}+H_{QPCi}\right)+E_\mathrm{C}\left(\frac{\hat{Q}}{e}-\frac{Q_0}{e}\right)^2,\label{Hcircuit}
\end{equation}
with $H_I$ ($H_{Li}$) the Hamiltonian describing the electron continuum in the island (in the lead $i$), $\hat{Q}$ the island charge operator, $Q_0$ a gate voltage dependent charge offset, and $H_{QPCi}$ the Hamitonian describing the electron transfers between island and lead $i$ (across QPC$_i$):
\begin{equation}
H_{QPCi}=t_i\sum_{k,k'}c_{Iik'}^+c_{Lik}^{\phantom{+}}+H.c.,\label{Hqpci}
\end{equation}
where $c_{Iik'(Lik)}$ is the electron annihilation operator in the island (lead) associated with the conduction channel across QPC$_i$ and $t_i$ the coupling coefficient (here assumed independent of $Lik$ and $Iik'$).
Introducing the electrons' `localization pseudospin-$1/2$' ($s_i$) between island (pseudospin state $\downarrow$) and lead $i$ (pseudospin state $\uparrow$), and regarding the island's charge $\hat{Q}$ as an independent macroscopic quantum degree of freedom (which is valid in the continuous density of states limit, see e.g. [32] for a specific discussion), the circuit Hamiltonian reads (see [20,29,32] and also \cite{Bao2017}):
\begin{eqnarray}
H=&&\,H_I+\sum_i\left[H_{Li}+ N_et_i(S^+s^-_i + S^-s^+_i)\right]\nonumber\\
&&+\,E_\mathrm{C}\left(\frac{\hat{Q}}{e}-\frac{Q_0}{e}\right)^2,\label{HcircuitNCK}
\end{eqnarray}
with $S^{\pm}=\sum_Q\ket{Q\pm1}\bra{Q}$, $s_i^+=(1/N_e)\sum_{k,k'}c_{Lik}^+c_{Iik'}^{\phantom{+}}$, $N_e\equiv\sqrt{\sum_{k,k'}}$ the effective number of electrons, and $s_i^-=(s_i^+)^\dagger$.
For small enough coupling coefficients $t_i$, the term proportional to $E_\mathrm{C}$ in eqs~\ref{Hcircuit} and \ref{HcircuitNCK} effectively freezes out all island's charge states in the limit $T\ll E_\mathrm{C}/k_\mathrm{B}$, except for the two of lowest energy (tuned at degeneracy in the Kondo regime).
In this low-temperature limit, the charge states of the island therefore reduce to a pseudospin $S$ of $1/2$, which is flipped by tunneling onto or off the island at each of the QPCs.
The terms of the form $S^+s^-_i$ in Eq.~\ref{HcircuitNCK} indeed describes a spin exchange between the Kondo impurity charge pseudospin-$1/2$ and the localization pseudospin-$1/2$ of the electrons in continuum $i$.
Note that the Kondo exchange is here proportional to $t_i$, in contrast with spin Kondo devices based on small quantum dots where spin flips result from virtual, second-order processes. 
Furthermore, the last term in Eq.~\ref{HcircuitNCK} reduces to a Zeeman splitting $\Delta E$ of the charge pseudospin of $1/2$, with an effective magnetic field that is simply proportional to the gate voltage detuning $\delta V_\mathrm{g}$ from charge degeneracy ($\Delta E=2E_\mathrm{C}|\delta V_\mathrm{g}|/\Delta$, for $|\delta V_\mathrm{g}|<\Delta/2$ with $\Delta$ the gate voltage period of Coulomb oscillations) [29].
Note that the corresponds to the anisotropic Kondo model, since there is no component $S^zs_i^z$ (in contrast to Eq.~1, where the coupling coefficient $J$ is assumed identical for the $x$, $y$ and $z$ components).
This anisotropy of the coupling is irrelevant in the renormalization group sense as it disappears in the low-temperature limit [13].
At $T\sim E_\mathrm{C}/k_\mathrm{B}$, the thermal activation of additional charge states effectively breaks the charge pseudospin-$1/2$ mapping, and thereby provides a high energy cutoff for Kondo physics (for a specific NRG study, see section ``charge' Kondo universality' and Fig.~S4).
Note finally that the experimentally probed electrical conductances across the charge Kondo electronic channels are connected, in usual Kondo language, to two-particle correlation functions (see [34,35] and also \cite{Yi2002,Pustilnik2017,Bao2017}).
Those correlation functions were calculated exactly at the fixed point by Ludwig and Affleck using conformal field theory [14] and, for the 2CK model, as a full function of temperature via NRG in [20].
In contrast, for previous spin 2CK nanostructures based on small quantum dots [2,5] the conductance is proportional to the single-particle $\mathcal{T}$ matrix [16].\\

{\noindent\textbf{Multichannel Kondo physics with strongly coupled contacts.}}
Beyond weakly coupled QPCs, quantum fluctuations compete with the freezing of higher energy charge states, which can break the direct mapping of the two lowest energy charge states on a Kondo pseudospin $S=1/2$ (note that charge quantization breakdown is found and predicted only close to the ballistic limit, see [31] for an experimental exploration).
Nevertheless, even for nearly ballistic contacts, where many charge states coexist in a quantum superposition, the low-temperature physics at the degeneracy point ($\delta V_g=0$) as well as the crossover flow from quantum criticality should be universal, indistinguishable from what one expects from the standard $S=1/2$ multichannel Kondo model.
This was shown in [15], in particular by mapping the circuit Hamiltonian derived for two nearly ballistic channels onto the Hamiltonian obtained by Emery and Kivelson [7] for the anisotropic two-channel Kondo model in the Toulouse limit (see below Eq.~53 of [15]).
It is also attested by the identical expression for the universal conductance flow $\tilde{G}_\mathrm{2CK}(T/T_\mathrm{co})$ for the crossover from 2CK quantum criticality (given by Eq.~\ref{G13FM} with $T/T_\mathrm{K}=0$) derived analytically in both the opposite tunnel and nearly ballistic limits (Eq.~6 in [20] and Eq.~38 in [30], respectively).
Besides these limits, the general case of arbitrary coupling strength was studied numerically (see e.g. [20,32] and section ``charge' Kondo universality'), further corroborating the robust Kondo character including in the presence of strongly coupled contacts.\\

{\noindent\textbf{Non-Fermi liquid temperature exponents.}} 
For $N\geq2$ Kondo channels and a Kondo impurity of spin $S=1/2$, observables are naively expected to display the dominant temperature power-law $T^{2/(N+2)}$ in the vicinity of the $N$CK fixed point (see e.g. [13] and references therein).
This is however not necessarily the case: different power laws can arise e.g. for observable involving operators in the charge or spin sectors [13] or due to a vanishing first order development near the fixed point.

For 2CK ($N=2$), the Kondo channels conductance in the present `charge' implementation is predicted to scale linearly ($\Delta G\propto T$), and not as the naively expected $\sqrt{T}$.
This prediction was obtained both analytically in the case of a nearly ballistic channel (see Eq.~A9 in [30], the specific discussion in Methods of [6], and Eq.~\ref{G13FM} with $T_\mathrm{co}=0$), and from recent NRG calculations starting from the opposite limit of a tunnel contact [20].
Note the difference with the conductance in previous spin 2CK implementations with small quantum dots, where the naively expected $\sqrt{T}$ was predicted and observed [2,5].
Indeed, the conductance in these spin Kondo devices is proportional to the single-particle $\mathcal{T}$ matrix [16], whereas in the `charge' implementation it relates to two-particle correlation functions (see [35] and also \cite{Bao2017,Pustilnik2017}, or Eq.~\ref{eq:kuboG}).
After submission of this manuscript, three new works calculating the 2CK power law dependence of the conductance in the `charge' Kondo implementation with different methods appeared (see [35] and also \cite{Bao2017,Pustilnik2017}): a linear behavior is also found in [35] and \cite{Bao2017,Pustilnik2017} (a different $T^2$ scaling initially obtained in the published article \cite{Pustilnik2017} was subsequently corrected in an Erratum and also in the second arXiv version).

For 3CK ($N=3$), the naively expected $T^{2/5}$ is precisely reproduced by the new NRG calculation of the universal `charge' Kondo conductance curve shown Fig.~4B,C.
This is best seen by plotting in a $\log$-$\log$ scale the difference $\Delta G$ between this NRG curve and the predicted 3CK fixed point $G_\mathrm{3CK}=2\sin^2(\pi/5)e^2/h$, as shown Fig.~S5.
Note that the new preprint \cite{Bao2017} finds a different power law $T^{4/5}$.
However, the general prediction $T^{4/(N+2)}$ of \cite{Bao2017} for arbitrary $N$ also seems in contradiction with previous calculations in the large $N$ limit of the frequency dependence at zero temperature $\omega^{2/(N+2)}$ (see eqs D30 and D11 in appendix D of \cite{Yi2002}), which instead correspond to the naively expected power law also predicted here by NRG for $N=3$.
In practice, we chose in the manuscript to keep using the NRG prediction $T^{2/5}$, that precisely matches the naive expectation.
From an experimental stand point, although the data is closer to $T^{2/5}$, it does not allow to unambiguously rule out $T^{4/5}$ due to the relatively large uncertainty combined with the possibility that observing $T^{4/5}$ requires approaching even closer of the fixed point.\\

{\noindent\textbf{$\mathbb{Z}_N$ parafermions.}} 
As a guide to the reader, we point out the particularly accessible discussion in the last paragraph of \cite{Bao2017}, which complements the original works [7,9].
The authors of \cite{Bao2017} relate the non-Fermi liquid character of the $N$-channel Kondo fixed point ($N\geq2$) to the $\mathbb{Z}_N$ parafermionic theory describing the renormalization flow at the generalized Toulouse point.
The parafermion charge $e\times(2N-2)/(N+2)$ is mentioned and a connection is made with the zero temperature residual entropy at the NCK fixed point ($S=\ln\left(2\cos\left[\pi/(N+2)\right]\right)$, as indicated in e.g. [13] and as also explicitly confirmed in the present context by our NRG calculations).\\
  
{\noindent\textbf{2CK conductance in near ballistic, low temperature limit  ($k_\mathrm{B}T\ll E_\mathrm{C}$, $1-\tau_{1,3}\ll1$).}} 
Here, we provide the theoretical expression used to calculate the conductance displayed as a continuous line in Fig.~1C and the universal crossover curve $\tilde{G}_\mathrm{2CK}(T/T_\mathrm{co})$ shown as a thick dashed line in Fig.~5B.
In addition, one can derive from this formula the quantitative theoretical expression of $T_\mathrm{co}$ for the crossover from 2CK quantum criticality for arbitrary $\delta V_\mathrm{g}$ at $1-\tau\ll1$ (used in Fig.~5B), and also the 2CK non-Fermi liquid scaling exponent for the conductance $\Delta G\propto T/T_\mathrm{K}$ at $\delta V_\mathrm{g}=0$ and $\tau_1=\tau_3\equiv\tau$ (Fig.~3), together with the tested functional form $T_\mathrm{K}\sim E_\mathrm{C}/(1-\tau)$ (bottom right inset of Fig.~3).
The analytical prediction of eqs~38, 26 and A9 in [30] give for the overall conductance $g_{1\text{-}3}$ across the two connected QPC$_{1,3}$ both set to be nearly ballistic, but not necessarily to identical tunings, and for low temperatures $T\ll E_\mathrm{C}/k_\mathrm{B}$: 
\begin{eqnarray}
g_{1\text{-}3} &=& 1/(1/G_1+1/G_3) \nonumber\\
&=& \frac{e^2}{2h} \Big[ 1- \frac{T}{T_\mathrm{K}^\star}  -\int^{\infty}_0 \frac{\cosh^{-2}(x)}{1+(2xT/T_\mathrm{co})^2}\mathrm{d}x \Big], \label{G13FM}
\end{eqnarray}
with
\begin{equation}
T_\mathrm{K}^\star =\frac{16 E_\mathrm{C}/\big(k_\mathrm{B}\pi^3 \exp(\mathbf{C}) \big)}{  2-\tau_1-\tau_3 + 2 \sqrt{(1-\tau_1)(1-\tau_3)} \cos(2 \pi \delta V_g/\Delta)  } ,\label{TKstarFM}
\end{equation}
\begin{eqnarray}
T_\mathrm{co} =&&\,\frac{2 \exp(\mathbf{C}) E_\mathrm{C}}{\pi^2k_\mathrm{B}}\times\\
&&\big( 2-\tau_1-\tau_3 - 2 \sqrt{(1-\tau_1)(1-\tau_3)} \cos(2 \pi \delta V_g/\Delta) \big),\nonumber
\end{eqnarray}
where $\mathbf{C}\simeq0.5772$ is the Euler-Mascheroni constant and $\Delta$ the gate voltage period of Coulomb oscillations.
Note that the numerical prefactor for the scaling Kondo temperature $T_\mathrm{K}^\star$ in Eq.~\ref{TKstarFM} does not precisely correspond to the convention used in the main article, which is based on the knowledge of the full renormalization flow only accessible through NRG methods (see next section for further discussion).
For two symmetric channels $\tau\equiv\tau_1=\tau_3$ and at charge degeneracy $\delta V_\mathrm{g}=0$, one finds $T_\mathrm{co}=0$ and $T_\mathrm{K}^\star\propto E_\mathrm{C}/(1-\tau)$.
The low temperature criteria $T\ll E_\mathrm{C}/k_\mathrm{B}$ therefore implies that Eq.~\ref{G13FM} is valid only in the asymptotic regime $T\ll T_\mathrm{K}^\star$, where $\Delta G\propto T$.
The universal 2CK crossover curve for the conductance per channel $\tilde{G}_\mathrm{2CK}(T/T_\mathrm{co})$ (thick dashed line in Fig.~5B) corresponds to $2g_{1\text{-}3}$ as given by Eq.~\ref{G13FM} in the limit of negligibly small $T/T_\mathrm{K}^\star$.
Importantly, the universal character of $\tilde{G}_\mathrm{2CK}$ is attested by the fact that the exact same expression was obtained in the opposite limit of tunnel contacts $\tau_{1,3}\ll1$ (Eq.~6 in [20]).
For two symmetric channels $\tau\equiv\tau_1=\tau_3$ at arbitrary charge degeneracy $\delta V_\mathrm{g}$, the crossover temperature simplifies into:
\begin{equation}
T_\mathrm{co}= 8\exp(\mathbf{C})E_\mathrm{C}(1-\tau)\sin^2(\pi\delta V_g/\Delta)/(k_\mathrm{B}\pi^2),\label{TcoFMsym}
\end{equation}
which was used to plot the data versus $T_\mathrm{co}/T$ in the right side of Fig.~5B.
Note that the generically expected quadratic dependence of $T_\mathrm{co}$ with the parameter-space distance to the 2CK quantum critical point is recovered for both small $\delta V_\mathrm{g}\ll\Delta$ and small $\Delta\tau\equiv\tau_1-\tau_3\ll1-\langle\tau_{1,3}\rangle$.
The continuous line in Fig.~1C corresponds the conductance $G_{1,3}=2g_{1\text{-}3}$ calculated with eqs~\ref{G13FM}, \ref{TKstarFM}, \ref{TcoFMsym} using the separately characterized values $\tau_1=\tau_3=0.9$, $T=7.9$\,mK, $E_\mathrm{C}=k_\mathrm{B}\times0.3$\,K and $\Delta=0.7$\,mV, without any fit parameters.\\

{\noindent\textbf{Kondo temperature versus model parameters.}} 
Here, we discuss the relationship between scaling/Kondo temperature and microscopic model parameters $\tau$ and $E_\mathrm{C}$, with a specific focus on the existence of a power-law divergence at a critical setting $\tau=\tau_c$.
The first characteristic of the Kondo temperature $T_\mathrm{K}$ is that it corresponds to the scaling temperature in the universal renormalization flow regime.
With such a definition alone, the multiplicative factor of $T_\mathrm{K}(\tau,E_\mathrm{C})$ is an arbitrary constant value.
Following standard practice \cite{Goldhaber-Gordon1998b}, this factor is set such that $G_{1(,2),3}(T=T_\mathrm{K})=G_\mathrm{2CK(3CK)}/2$.
The Kondo temperature therefore also corresponds to the characteristic temperature scale at which Kondo physics develops when starting with a weak Kondo coupling (e.g. $\tau\ll1$).
Beyond initially weak Kondo couplings, $G_\mathrm{2CK(3CK)}/2$ may not be within the explored range of renormalized conductances  (e.g. for large $\tau$).
However, the scaling Kondo temperature $T_\mathrm{K}$ can always be adjusted by matching the data in the universality regime ($T\ll E_\mathrm{C}/k_\mathrm{B}$) with the full universal curve obtained by NRG starting from a weak tunnel coupling.
The definition of $T_\mathrm{K}$ as the scaling Kondo temperature allows for values possibly much larger than $E_\mathrm{C}/k_\mathrm{B}$.
In such cases, $T_\mathrm{K}$ does not correspond to the temperature scale at which Kondo physics develops since the universal Kondo regime only takes place well below the high-energy cutoff $E_\mathrm{C}$.
For 2CK, Matveev and Furusaki predict in [30] that $T_\mathrm{K}(1-\tau\ll1,E_\mathrm{C})\propto E_\mathrm{C}/(1-\tau)$, where $\tau_c=1$ (see previous section).
Note that a peaked Kondo temperature at a specific Kondo coupling setting was subsequently predicted for 2CK in [32,35].
In general (beyond 2CK), the power-law scaling of $T_\mathrm{K}$ at $|\tau-\tau_c|\ll1$ can be obtained assuming that the Kondo temperature $T_\mathrm{K}$ diverges at a critical transmission probability $\tau_c$ (in agreement with experimental observation, see insets in figs~3 and 4B).
Expanding linearly the channels conductance for $|\tau-\tau_c|\ll1$, one finds at $T\ll E_\mathrm{C}/k_\mathrm{B}$:
\begin{equation}
G_i(T)=G_\mathrm{NCK}\times\left( 1+A(\tau-\tau_c)(k_\mathrm{B}T/E_\mathrm{C})^\alpha \right),
\end{equation} 
with $A$ a multiplicative factor of order $1$, $G_\mathrm{NCK}$ the NCK low-temperature conductance fixed point, and $\alpha$ the temperature exponent for the conductance when approaching the Kondo fixed point ($\alpha=1$ for 2CK , $\alpha=2/5$ for 3CK).
Comparing with the low-temperature Kondo scaling $G_i(T)-G_\mathrm{NCK}\propto(T/T_\mathrm{K})^\alpha$, where all microscopic parameters are encapsulated in $T_\mathrm{K}$, directly gives
\begin{equation}
T_\mathrm{K}(|\tau-\tau_c|\ll1)\propto E_\mathrm{C}|\tau-\tau_c|^{-1/\alpha}.
\end{equation}
This prediction, which is novel for 3CK, is shown as continuous lines in the insets of Fig.~3 using the corresponding theoretical values of $\alpha$.
A close agreement is observed with the data.\\
  
{\noindent\textbf{Asymptotic crossover limit  $\tilde{G}_\mathrm{NCK}(T_\mathrm{co}/T\ll1)$.}} 
Here we derive the asymptotic functional form at $T_\mathrm{co}/T\rightarrow0$ (close to quantum criticality) of the universal conductance curve $\tilde{G}(T_\mathrm{co}/T)$ describing the crossover from NCK Kondo quantum criticality induced by breaking the degeneracy of the charge Kondo pseudospin ($\Delta E\neq0$, $\Delta\tau=0$, $T\ll T_\mathrm{K}$).
These asymptotic functional forms are shown for 2CK and 3CK as grey dash-dotted lines in the right side of Fig.~5B and Fig.~5C, respectively.
The limit $T_\mathrm{co}/T\rightarrow0$ corresponds to very small gate voltage detuning $\delta V_\mathrm{g}/\Delta\ll1$ ($\Delta$ being the gate voltage period).
Although the physics is dominated by non-Fermi liquid scalings induced by the NCK quantum critical point, the expansion at $T\neq0$ of the physical conductance observable with respect to $\delta V_\mathrm{g}$ is regular (analytic), as finite temperature regularizes infrared divergences in the corresponding coefficients.
From the even symmetry between positive and negative detunings $\delta V_\mathrm{g}$, the first term in this expansion is quadratic in $\delta V_\mathrm{g}$.
In the presently considered regime $T\ll T_\mathrm{K}$, this reads $\Delta G\equiv G(\delta V_\mathrm{g}\rightarrow0)-G_\mathrm{NCK}\propto\delta V_\mathrm{g}^2$.
On the other hand, we generally know that $G(T\ll T_\mathrm{K},\delta V_\mathrm{g})$ can be reduced to a universal function $\tilde{G}_\mathrm{NCK}$ of the rescaled temperature $T/T_\mathrm{co}$ (for $T_\mathrm{co}\ll T_\mathrm{K}$).
Moreover, using the analogy between Zeeman splitting of a magnetic impurity and charge pseudospin energy detuning ($\Delta E\propto\delta V_\mathrm{g}$), conformal field theory predicts that the crossover temperature scales as $T_\mathrm{co}\propto\delta V_\mathrm{g}^{(N+2)/N}$ in the limit of small $\delta V_\mathrm{g}$ (see e.g. [13]).
By direct identification, one immediately deduces the power-law asymptotic behavior $\tilde{G}_\mathrm{NCK}(T_\mathrm{co}/T\ll1)-G_\mathrm{NCK}\propto(T_\mathrm{co}/T)^{2N/(N+2)}$.
At 2CK, this expression reduces to a linear asymptotic scaling $\tilde{G}_\mathrm{2CK}(T_\mathrm{co}/T\ll1)-e^2/h\propto(T_\mathrm{co}/T)$, in agreement with the full analytical prediction (Eq.~\ref{G13FM}, see right side of Fig.~5B).
At 3CK, the above asymptotic expression reduces to $\tilde{G}_\mathrm{3CK}(T_\mathrm{co}/T\ll1)-G_\mathrm{3CK}\propto(T_\mathrm{co}/T)^{6/5}$, in agreement with novel NRG calculations (see right side of Fig.~5C).\\

{\noindent\textbf{Super-ballistic conductance.}}
NRG calculations directly show that the emergence of a super-ballistic single-channel conductance follows from the Kondo model.
It arises along the non-monotonous renormalization flow towards the 1CK fixed point, when considering the conductance observable in the present `charge' implementation.
Note that a similar observation (although less substantial) was previously made when investigating the `charge' Kondo renormalization flow with two channels (see Methods in [6] for a specific discussion).
Experimentally, the opening of a second channel across QPC$_2$ could, in principle, provide a simple explanation for the measurement of a conductance $G_2$ above $e^2/h$.
However this simple explanation can be directly ruled out, without the need to invoke the NRG confirmation.
Firstly, the second and third (inner) quantum Hall edge channels that could possibly be transmitted across QPC$_2$ (the experiment is performed at filling factor $\nu=3$) are completely reflected, by a large margin, when the first (outer) quantum Hall edge channel is partially transmitted.
This is evidenced by the very broad ($0.25$\,V in split gate voltage $V_2^\mathrm{qpc}$) and very flat $e^2/h$ plateau, which separates the full opening of the first channel from the point where the second channel starts to open (see inset of Fig.~S2E).
Secondly, this $e^2/h$ plateau is very robust up to energies much higher than the charging energy $E_\mathrm{C}\simeq26\,\mu e$V that sets an upper bound for Coulomb and Kondo effects:
We checked the plateau robustness up to a dc voltage of $V_\mathrm{dc}\simeq70\,\mu$V applied directly across QPC$_2$ (we did not try higher values), and found that the plateau remains very precisely at $G_2=e^2/h$.
Moreover, we find that a QPC initially completely closed stays closed even in the presence of the charge Kondo effect (here with two symmetric channels as shown with the $G_2=0$ data in Fig.~6, and also with two asymmetric channels in the previous `charge' Kondo experiment described in [6]).
In fact, the conductance of a channel that is initially (at high $T$) less coupled to the island than the other ones is found and predicted to be systematically further suppressed as temperature is reduced (e.g. an hypothetical weakly transmitted second channel across QPC$_2$, despite the above evidences that there is no such second channel).
In contrast, we observe here a large $G_2$ overshoot, of up to $+0.25e^2/h$ above the free electron quantum limit $e^2/h$.
We now list the specific QPC tunings of $\tau_{1,2,3}$ for which we found a super-ballistic conductance $G_2$ in Fig.~6 (in the explored temperature range, only including data points for which the statistical uncertainty on $G_2$ is smaller than $0.1e^2/h$, and with the discrete settings of $\tau_2$ in the indicated ranges as given Fig.~S2C):
at $\tau_{1,3}\simeq0.1$ for $\tau_{2}\in[0.79,0.94]$; at $\tau_{1,3}\simeq0.20$ for $\tau_{2}\in[0.64,0.98]$; at $\tau_{1,3}\simeq0.34$ for $\tau_{2}\in[0.68,0.98]$; at $\tau_{1,3}\simeq0.48$ for $\tau_{2}\in[0.74,0.98]$; at $\tau_{1,3}\simeq0.56$ for $\tau_{2}\in[0.79,0.98]$; at $\tau_{1,3}\simeq0.64$ for $\tau_{2}\in[0.82,0.98]$; at $\tau_{1,3}\simeq0.68$ for $\tau_{2}\in[0.85,0.98]$; at $\tau_{1,3}\in\{0.74,0.79\}$ for $\tau_{2}\in[0.90,0.98]$; at $\tau_{1,3}\in\{0.82,0.85\}$ for $\tau_{2}\in\{0.94,0.98\}$; at $\tau_{1,3}\in\{0.89,0.94\}$ for $\tau_{2}\simeq0.98$.

\subsection{Numerics}

\begin{figure*}[htb]
\renewcommand{\thefigure}{S\arabic{figure}}
\renewcommand{\figurename}{Fig.}
\centering\includegraphics [width=0.95\textwidth]{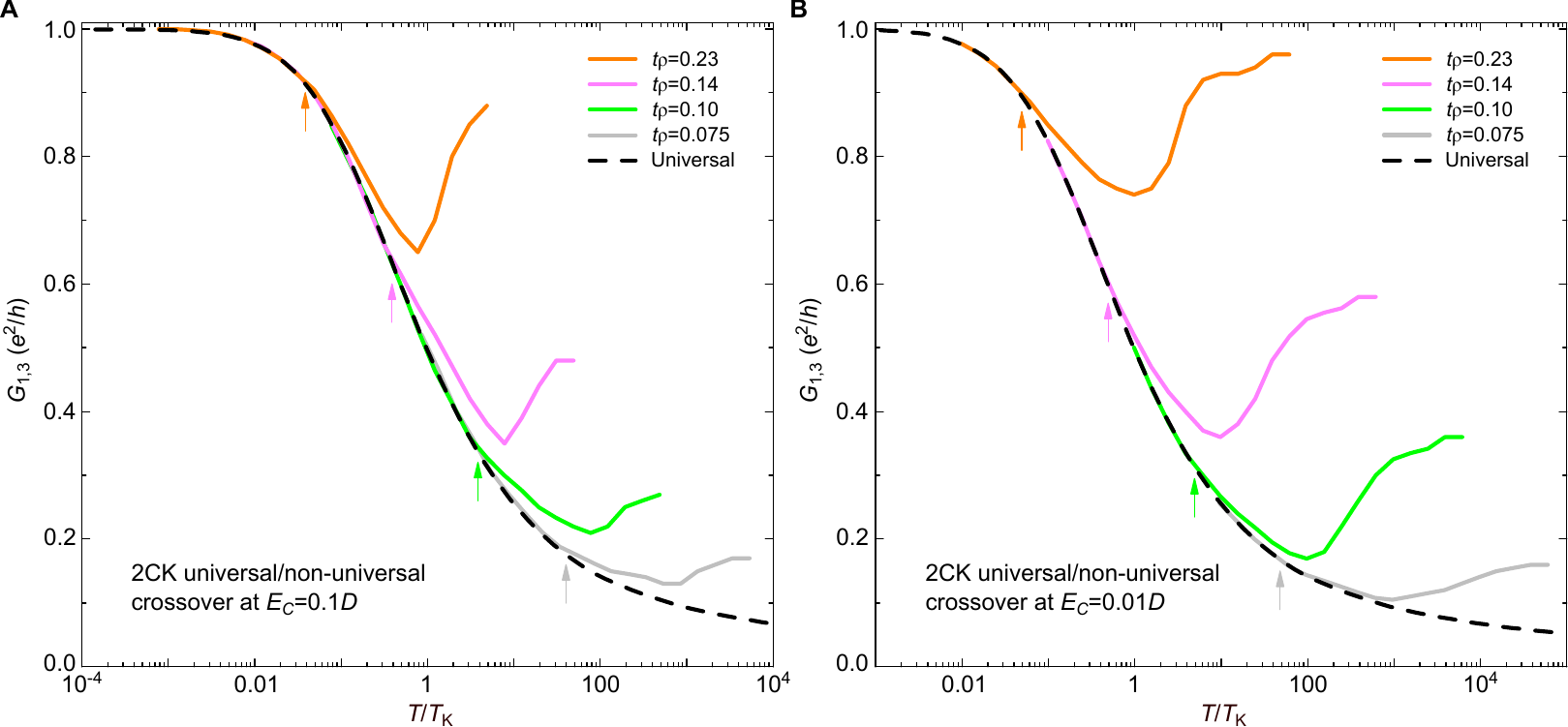}
\caption{\small
\textbf{Universality criterion} at finite $E_\mathrm{C}$ and arbitrary QPC couplings. 
NRG calculations of the 2CK channels conductance $G_{1,3}$ are plotted versus rescaled temperature $T/T_\mathrm{K}$.
The universal conductance curve shown as a black dashed line (calculated in the regime $k_\mathrm{B}T_\mathrm{K}\ll D\ll E_\mathrm{C}$, i.e. with two charge states and tunnel contacts) is compared to finite $E_\mathrm{C}$ calculations (colored continuous lines; $E_\mathrm{C}=0.1D$ in panel (A), $E_\mathrm{C}=0.01D$ in panel (B); including twenty charge states of the metallic island in NRG) for different settings of the normalized QPC coupling coefficient $t\rho$ (with $t$ defined Eqs.~\ref{Hqpci},\ref{HcircuitNCK}, and $\rho$ the `effective' electronic density of states).
The universality criterion $k_\mathrm{B}T_\mathrm{uni}\equiv E_\mathrm{C}/20$ is pointed out with vertical arrows of the same color as the corresponding NRG calculation at finite $E_\mathrm{C}$.
}
\end{figure*}

\begin{figure}[htb]
\renewcommand{\thefigure}{S\arabic{figure}}
\renewcommand{\figurename}{Fig.}
\centering\includegraphics [width=\columnwidth]{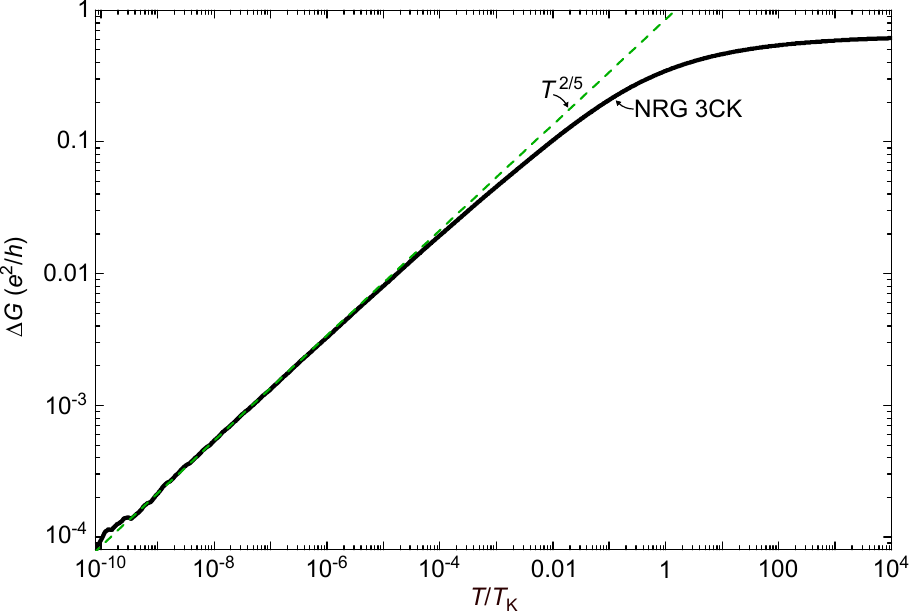}
\caption{\small
\textbf{3CK non-Fermi liquid power-law.}
The $T^{2/5}$ power-law (straight dashed line) naively expected near the 3CK fixed point is compared with the 3CK universal conductance curve calculated by NRG, as a function of the rescaled temperature $T/T_\mathrm{K}$ in a $\log$-$\log$ scale.
The black continuous line displays the difference $\Delta G$ between NRG calculation and predicted fixed point $G_\mathrm{3CK}=2\sin^2(\pi/5)e^2/h$.
}
\end{figure}

{\noindent\textbf{Numerical renormalization group calculations.}} 
Numerical calculations of the universal 2CK and 3CK conductance presented in figs~4,5,6 were performed using a variant of Wilson's numerical renormalization group (NRG) technique (see e.g. [18] and \cite{Wilson1975}), adapted to treat the multichannel charge-Kondo Hamiltonian, Eq.~\ref{Hcircuit}, in the limit where the island charging energy $E_\mathrm{C}$ is the largest energy scale in the problem (including the conduction electron half-bandwidth $D$; the effect of $E_\mathrm{C}<D$ is discussed in the next section). 
In this case, the two charge states of the island of lowest energy form a pseudospin-$1/2$, while the other charge states can be ignored whatever the QPCs' connection strengths.
The resulting anisotropic multichannel Kondo model (Eq.~\ref{HcircuitNCK} including only the two lower charge states) is solved non-perturbatively with NRG. 
The conduction electron density for each of the $N$ channels is discretized logarithmically and the system is then mapped onto a 1-dimensional model in which the Kondo pseudospin is connected to one end of the bundle of $N$ semi-infinite `Wilson chains'.
The renormalization group character of the problem is revealed by iterative diagonalization of the chain: the physics at successively lower energy scales is probed at each step as high-lying states are iteratively eliminated.
The computational complexity of an NRG calculation scales exponentially with the number of channels, $N$. 
Consequently, although standard NRG methods could still be used for 2CK (Fig.~4A,C), the charge-3CK model would have been essentially intractable (large symmetries are broken due to the spin anisotropy inherent to the charge-Kondo setup, and further reduced in the vicinity of the critical point by channel asymmetry and gate detuning of charge degeneracy).
This limitation was overcome by using for the 3CK calculations (shown figs~4B,C, 5C and 6) the recently-developed `interleaved NRG' (iNRG), which makes use of a modified discretization to combine the $N$ Wilson chains into a single generalized chain \cite{Mitchell2014,Stadler2016}.
The experimental quantity of interest is the linear response dc differential conductance, whose accurate calculation requires further modification of the standard NRG procedure, as now briefly described. 
To the Hamiltonian Eq.~\ref{Hcircuit} we add a time-dependent bias term to lead $i$, $H_{\text{bias}}=e V_i\hat{N}_{Li}\cos(\omega t)$, where $\hat{N}_{Li}=\sum_k c_{Lik}^{+}c_{Lik}^{\phantom{+}}$ is the total electron number operator for lead $i$. 
Measurement of the resulting current into lead $j$ allows determination of elements of the conductance tensor $G_{ij}(T,V_i)=d\langle \hat{I}_{j}\rangle_T/d V_i$.
Within linear response $V_i\rightarrow 0$, we employ the Kubo formula \cite{Izumida1997},
\begin{equation}
\label{eq:kuboG}
G_{ij}(T)=\frac{e^2}{h}\lim_{\omega\rightarrow0}\left[\frac{2\pi\hbar^2\,\text{Im}\,K_{ij}(\omega,T)}{\hbar\omega}\right],
\end{equation}
where the limit $\omega\rightarrow 0$ yields the desired dc conductance.
$K_{ij}(\omega,T)$ is the Fourier transform of the retarded current-current correlator, $K_{ij}(t,T)=\mathrm{i}\theta(t) \langle [\dot{N}_{Lj}(t),\dot{N}_{Li}(0) ] \rangle$, where $\dot{N}_{L\alpha}=\tfrac{d}{dt} \hat{N}_{L\alpha}$. 
In NRG, $K_{ij}(\omega,T)$ is obtained directly on the real axis as an entire function of $\omega$ for any $T$.
It is calculated using the full density matrix, established on a complete basis \cite{Weichselbaum2007}.
However, straight application of this approach is plagued by numerical instabilities: $K_{ij}(\omega,T)$ must be determined very accurately to avoid the spurious divergence of $K(\omega,T)/\omega$ on taking the $\omega\rightarrow 0$ limit.
Instead, we exploit the identity $\text{Im}\,K_{ij}(\omega,T)/\omega=\omega \text{Im}\,\tilde{K}_{ij}(\omega,T)$, which we derived from equations of motion, where $\tilde{K}_{ij}(t,T)=\mathrm{i}\theta(t) \langle [\hat{N}_{Lj}(t),\hat{N}_{Li}(0) ] \rangle$.
This trick is found to drastically improve the accuracy of the conductance calculation in NRG.
It is especially important in the three channel case, whose fixed point conductance takes a nontrivial intermediate value (see [34] and also \cite{Yi2002,Bao2017}).
The true `universal' renormalization flow of the conductance is obtained formally in the scaling limit $k_\mathrm{B}T_K/D\rightarrow 0$. 
In practice, we use $\rho t=0.025$ ($\rho\equiv N_e/2D$, $t\equiv t_{1(,2),3}$) yielding $T_K\sim 10^{-15}D/k_\mathrm{B}$.
For 2CK NRG calculations, we used a discretization parameter \cite{Wilson1975} $\Lambda=3$, retained $N_s=10000$ states at each step, and averaged the results of $N_z=4$ calculations.
For 3CK iNRG calculations, we used $\Lambda=3$, $N_s=38000$, and $N_z=3$.
The three spinful channels, each with U(1) conserved charge, were interleaved, and global U(1) spin symmetry was exploited (note the inherent spin-anisotropy of the charge-Kondo setup).\\

\noindent\textbf{`Charge' Kondo universality.}
To what extent does one recover universal Kondo physics at low temperature in the present `charge' implementation: 
Does it depend on the channels coupling strength or on the ratio $E_\mathrm{C}/D$?
How small does the temperature need to be in practice?
We systematically find, based on 2CK NRG calculations, that the universal Kondo scaling curve for the conductance is accurately recovered for $T\lesssim \min(E_\mathrm{C},D)/20k_\mathrm{B}$, including when the deduced scaling temperature $T_\mathrm{K}$ is large with respect to $E_\mathrm{C}/k_\mathrm{B}$ or $D/k_\mathrm{B}$.
Previous calculations in [20] investigated the non-universal behavior resulting from the finite conduction electron bandwidth $D$, which was assumed to be much smaller than $E_\mathrm{C}$.
In that case, whatever the temperature and coupling strengths, only two charge states are accessible and need to be included in the calculation.
Here we consider the effect of a finite island charging energy $E_\mathrm{C}$ in Eq.~\ref{Hcircuit}, which requires going beyond the spin-$\frac{1}{2}$ Kondo paradigm.
In practice, 20 charge states are taken into account in the calculations.
In Fig.~S4, the full universal 2CK conductance scaling curve (black dashed line; two charge states, tunnel contacts) is plotted versus $T/T_\mathrm{K}$ and compared to NRG calculations at finite $E_\mathrm{C}$ (colored continuous lines; $E_\mathrm{C}=0.1D$ in panel A, $E_\mathrm{C}=0.01D$ in panel B) for different values of the normalized QPCs coupling $\rho t\in \{0.075,0.10,0.14,0.23\}$, with $\rho$ the electronic density of states per unit energy ($\rho\equiv N_e/2D$) and $t$ defined eqs~\ref{Hqpci},\ref{HcircuitNCK}. 
The identical scaling observed for all values of $t$ at low enough $T/T_\mathrm{K}$, systematically seen also in the opposite regime $D\ll E_\mathrm{C}$ (not shown), shows that there exists a temperature $T_\mathrm{uni}(t,D,E_\mathrm{C})$ below which the measured channels conductance $G_i(T)$ collapses to the same universal Kondo curve when rescaled by $T_K$ -- independently of $T_\mathrm{K}/D$, $T_\mathrm{K}/E_\mathrm{C}$, $E_\mathrm{C}/D$ or $\rho t$. 
Defining $T_{\mathrm{uni}}$ as the temperature below which deviations from universality are smaller than 0.01$e^2/h$, we find numerically $k_\mathrm{B} T_{\mathrm{uni}}\sim E_\mathrm{C}/20$ for $E_\mathrm{C}<D$, and more generally $k_\mathrm{B} T_{\mathrm{uni}}\sim \min(E_\mathrm{C},D)/20$.
Vertical arrows in Fig.~S4 indicate the position of $E_\mathrm{C}/20k_\mathrm{B}$ in rescaled temperature, with the same color code as the corresponding finite $E_\mathrm{C}$ NRG calculation.
Note that the pronounced deviations from universality at higher temperature strongly depend on the ratio $E_\mathrm{C}/D$ as well as on the number of charge states included in the calculation. 
While we naively expect a similar universality criterion for 3CK, it should be noted that in this case two different universal curves exist on approaching from below ($\tau <\tau_c$) or above ($\tau >\tau_c$) because the critical 3CK fixed point conductance takes an intermediate value ($0<G_{\mathrm{3CK}}<e^2/h$).\\

\noindent\textbf{3CK-Fermi liquid crossover.}
We address the crossover from 3CK quantum criticality induced by an energy splitting $\Delta E\propto\delta V_\mathrm{g}$ of the `charge' Kondo impurity.
Is the universal character of the crossover curve $\tilde{G}_\mathrm{3CK}(T_\mathrm{co}/T)$ preserved over the full range of gate voltage detuning $\delta V_\mathrm{g}$, like at 2CK (for 2CK see Eq.~\ref{G13FM}, derived for near ballistic channels such that $T_\mathrm{K}\gg E_\mathrm{C}/k_\mathrm{B}\gg T_\mathrm{co}$)?
If it is the case, what is the generalized, periodic expression of $T_\mathrm{co}$ versus arbitrary detuning $\delta V_\mathrm{g}$ (for 2CK, see Eq.~\ref{TcoFMsym})?
Here, we detail NRG calculations that establish the experimental observation of a robust universality for $\tilde{G}_\mathrm{3CK}(T_\mathrm{co}/T)$, for any gate voltage detuning, as well as the generalized expression for the crossover temperature $T_\mathrm{co}\propto\sin^{5/3}(\pi\delta V_\mathrm{g}/\Delta)$ (see Fig.~5C).
First, the universal form $\tilde{G}_\mathrm{3CK}(T/T_\mathrm{co})$ shown in Fig.~5C was obtained in the standard Kondo limit (retaining just two charge states, and with $T_\mathrm{co}\ll T_\mathrm{K}$).
Second, we focus on the more challenging numerical study of the relationship between $T_\mathrm{co}$ and arbitrary $\delta V_{\rm g}$.
This requires to directly calculate the correct form of the full Coulomb peaks (see Fig.1 and Fig.~5B,C), for which one must simulate the effect of finite $\delta V_{\rm g}$ over an entire charging period. 
This involves including many charge states in the calculation (as in the previous section), beyond the standard Kondo model description restricted to the two lower charge states.
From analysis of the temperature-dependence of the conductance for a given gate voltage $\delta V_{\rm g}$ within these NRG calculations, the crossover scale $T_\mathrm{co}(\delta V_{\rm g})$ could be extracted directly.
For 3CK, we find a robust power law $T_\mathrm{co}\propto\delta V_{\rm g}^{5/3}$ for small $\delta V_{\rm g}$, as expected from conformal field theory.
But for larger $\delta V_\mathrm{g}$, we find marked deviations, which follow the generalized periodic form $T_\mathrm{co}\propto\sin^{5/3}(\pi\delta V_{\rm g}/\Delta)$, to within a numerical accuracy better than $0.5\%$. 
This is reminiscent of the periodic variation of the crossover scale in the 2CK model near perfect transmission, as derived analytically by Matveev.
We note, however, that no such analytical predictions as yet exist for 3CK, since the critical point is irreducibly interacting.
Interestingly, the data (both experimental and numerical) over the entire range of $\delta V_{\rm g}$, when rescaled in terms of this crossover scale $T_\mathrm{co}$, fit to the universal form $\tilde{G}_\mathrm{3CK}(T/T_\mathrm{co})$ -- analogous to the behaviour in the 2CK case. 
This novel finding extends and strengthens the notion of universality in the 3CK system.\\
